\documentclass[sn-mathphys-num]{sn-jnl}

\usepackage{graphicx}%
\usepackage{multirow}%
\usepackage{amsmath,amssymb,amsfonts}%
\usepackage{amsthm}%
\usepackage{mathrsfs}%
\usepackage[title]{appendix}%
\usepackage{xcolor}%
\usepackage{textcomp}%
\usepackage{manyfoot}%
\usepackage{booktabs}%
\usepackage{algorithm}%
\usepackage{algorithmicx}%
\usepackage{algpseudocode}%
\usepackage{listings}%

\usepackage{booktabs} 

\theoremstyle{thmstyleone}%
\theoremstyle{thmstyletwo}%

\theoremstyle{thmstylethree}%

\begin{document}

\title[Article Title]{How to quantify an examination? Evidence from physics examinations via complex networks}

\author[1]{\fnm{Min} \sur{Xia}}

\author*[2]{\fnm{Zhu} \sur{Su}}\email{suz@mail.ccnu.edu.cn}

\author*[1]{\fnm{Wei-Bing} \sur{Deng}}\email{wdeng@mail.ccnu.edu.cn}

\author*[3]{\fnm{Xiu-Mei} \sur{Feng}}\email{xiumeifeng@mail.ccnu.edu.cn}


\author*[1]{\fnm{Ben-Wei} \sur{Zhang}}\email{bwzhang@mail.ccnu.edu.cn}

\affil[1]{\orgname{Key Laboratory of Quark \& Lepton Physics (MOE) and Institute of Particle Physics, Central China Normal University}, \city{Wuhan}, \postcode{430079}, \country{China}}

\affil[2]{\orgname{National Engineering Laboratory for Educational Big Data, Central China Normal University}, \city{Wuhan}, \postcode{430079}, \country{China}}

\affil[3]{\orgname{College of Physical Science and Technology, Central China Normal University}, \city{Wuhan}, \postcode{430079}, \country{China}}

\abstract{Given the untapped potential for continuous improvement of examinations, quantitative investigations of examinations could guide efforts to considerably improve learning efficiency and evaluation and thus greatly help both learners and educators. However, there is a general lack of quantitative methods for investigating examinations. To address this gap, we propose a new metric via complex networks; i.e., the knowledge point network (KPN) of an examination is constructed by representing the knowledge points (concepts, laws, etc.) as nodes and adding links when these points appear in the same question. Then, the topological quantities of KPNs, such as degree, centrality, and community, can be employed to systematically explore the structural properties and evolution of examinations. In this work, 35 physics examinations from the Chinese National College Entrance Examination spanning from 2006 to 2020 were investigated as an evidence. We found that the constructed KPNs are scale-free networks that show strong assortativity and small-world effects in most cases. The communities within the KPNs are obvious, and the key nodes are mainly related to mechanics and electromagnetism. Different question types are related to specific knowledge points, leading to noticeable structural variations in KPNs. Moreover, changes in the KPN topology between examinations administered in different years may offer insights guiding college entrance examination reforms. Based on topological quantities such as the average degree, network density, average clustering coefficient, and network transitivity, the comprehensive difficulty coefficient is proposed to evaluate examination difficulty. All the above results show that our approach can objectively and comprehensively quantify the knowledge structures and examination characteristics. These networks may elucidate comprehensive examination knowledge graphs for educators and guide adjustments and improvements in teaching methodologies.}




\maketitle

\section{Introduction}\label{sec1}
Examination, known for its scientific and impartial nature, is frequently employed to evaluate learning efficacy and teaching effectiveness \cite{brookhart2009assessment}. Some large-scale examinations, often referred to as high-stakes tests, are commonly utilized for measuring school effectiveness and determining students' advancement in education and employment \cite{au2007high, jacob2005accountability, von2016influence, fang2022does}. The scientific and rational design of examinations influences the fairness and effectiveness of assessments \cite{mccoubrie2004improving}. Consequently, evaluating the examinations has important value and great significance, helping us to achieve a comprehensive understanding of the alignment between examination objectives and teaching practices, revealing the inherent patterns of examinations and educational effectiveness \cite{mulbar2017analysis}, and providing essential reference points for educational reforms and the enhancement of teaching quality \cite{hallinger2014teacher}. However, how to quantify the examinations remains a considerable challenge. On the one hand, current methods are often subjective, biased by personal experience, and limited in scope \cite{garg2013analytical}. It is also difficult to track changes over time through methods that compare syllabuses or exam standards \cite{han2024alignment,akhtar2020measurement}. On the other, while there has been a shift from initial expert judgments towards data-driven approaches like multiple regression analysis \cite{rupp2001combining} and neural networks \cite{perkins1995predicting}, these approaches still require manual labeling, which are technically complex, require large sample sizes, and exhibit poor interpretability of results \cite{bi2024difficulty}.

Complex networks may help us to address the challenge of quantifying examinations. Networks are useful for representing large and complex datasets as clear hierarchies \cite{jiang2014structure} and offering a fundamental framework for analyzing the dynamic evolution mechanisms of complex systems \cite{sun2020revealing}. A notable advantage of networks is their ability to reveal the interactions and scaling behavior among various components. Numerous everyday scenarios can be conceptualized as complex networks \cite{fox2020intrinsic}. Therefore, complex networks have been applied to study structures and correlations within complex systems in various fields. In fact, complex networks are also widely used in education, such as curriculum design \cite{forsman2014extending}, teacher-student interactions \cite{lu2020diversities, wolf2021complex}, classroom teaching \cite{goh2014complex,shu2018determining}, citation coauthorship networks \cite{feng2020mixing}, learning pathways \cite{ramirez2019students}, and textbook analysis \cite{yun2018extraction,stella2017multiplex, sizemore2018knowledge}. Examinations test students’ mastery of knowledge by setting a series of questions, each corresponding to one or more knowledge points. If knowledge points are viewed as nodes, and their relationships are viewed as edges, then the complex knowledge system formed by each examination can be represented by a network. Thus, we can employ complex networks to quantitatively analyze the examinations. Nonetheless, to our knowledge, few studies have applied complex networks to examination analysis. Given their advantages in terms of scale, duration, and multidimensionality, the application of complex networks to examination analysis will provide us with new insights.

When discussing examinations, one must consider the Chinese National College Entrance Examination (NCEE), commonly known as the "Gaokao." As the largest examination in China and globally, the Gaokao plays a crucial role in ensuring the fairness and accessibility of Chinese education, profoundly influencing the development of education and talent cultivation \cite{zhang2016national}. For more than a decade, nearly 10 million Chinese students have taken the Gaokao every year. This underscores the importance of exam question design. Utilizing the NCEE as research material enhances the sample's validity and generalizability, making the conclusions more practically significant and broadly applicable. Based on this, this article examines 35 physics examination papers from NCEE (spanning from 2006 to 2020) and employs network analysis to conduct a case study, aiming to provide a new method and tool for examination analysis and to achieve comprehensive quantitative evaluation of examinations.

Specifically, we constructed a network model to describe the knowledge structures of college entrance examinations, abstracted the knowledge content of test questions, and constructed a knowledge point network (KPN) on the basis of the knowledge point distribution across different questions. By conducting network analysis on the KPNs derived from the 35 physics examination papers, we explored the underlying topological characteristics and the significance of such networks for educators and learners. In addition, we considered statistical models and the evolutionary characteristics of the examinations. Moreover, we explored the change in the examination difficulty using network indicators derived from the KPNs, and the results were validated with various network indicators, such as the density and average clustering coefficient.

It was found that the constructed KPNs are scale-free networks that show strong assortativity and small-world effects in most cases. The communities within the KPNs are obvious, and the key nodes are mainly related to mechanics and electromagnetism. Changes in the KPN topology between examinations administered in different years may offer insights guiding college entrance examination reforms. The characteristics reflected by KPNs of different volumes align with the intended design of the NCEE. Different question types are related to specific knowledge points, leading to noticeable structural variations in KPNs. Furthermore, our study also verified from various perspectives that the $F_d$ constructed based on four indicators including average degree, network density, average clustering coefficient, and network transitivity can effectively reflect examination difficulty.

These findings demonstrate that our approach can objectively and comprehensively quantify the knowledge structure and characteristics of examinations. These networks enhance educators’ understanding of comprehensive examination knowledge graphs, providing a new perspective for exam analysis, teaching adjustments, and improvements. Our method can be applied to assess exam difficulty in test design, analyse and provide feedback on exams post-assessment, and guide teachers in daily instruction based on the network properties of examinations, thereby holding promise for application in educational practice.

\section{Results}\label{sec2}

\subsection{Topological Structure of KPNs}\label{subsec1}

\begin{table}[h]
	\caption{\textbf{Main statistical characteristics of various knowledge point networks (KPNs).} This table shows the results of the IKPN, $\left\langle \text{KPN} \right\rangle$, and four randomly selected KPNs. IKPN represents the network integrating all the KPNs, $\left\langle \text{KPN} \right\rangle$ represents the result of averaging all the KPNs. Different volumes in various years are abbreviated; for example, 06V1 represents the KPN of National Volume 1 from 2006. \# of nodes represents the number of nodes in the KPN. \# of edges represents the number of edges in the KPN.}\label{tab1}%
	\begin{tabular}{@{}lllllll@{}}
		\toprule
		Topological  quantities        & IKPN  & $\left\langle \text{KPN} \right\rangle$ & 06V1  & 13V2  & 17V1  & 20V2  \\ 
		\midrule
		\# of nodes                    & 336   & 43                                      & 42    & 46    & 42    & 45    \\
		\# of edges                    & 1559  & 72                                      & 70    & 72    & 69    & 72    \\
		Diameter                       & 8     & 4                                       & 3     & 4     & 3     & 4     \\
		Density                        & 0.028 & 0.075                                   & 0.081 & 0.070 & 0.080 & 0.073 \\
		Assortativity                  & 0.002 & 0.393                                   & 0.466 & 0.081 & 0.551 & 0.428 \\
		Transitivity                   & 0.32  & 0.78                                    & 0.86  & 0.63  & 0.75  & 0.73  \\
		Average degree                 & 9.28  & 3.16                                    & 3.33  & 3.13  & 3.29  & 3.20  \\
		Average shortest path length   & 3.62  & 2.00                                    & 1.95  & 2.28  & 2.00  & 2.21  \\
		Average clustering coefficient & 0.04  & 0.55                                    & 0.85  & 0.38  & 0.72  & 0.37  \\
		\bottomrule
	\end{tabular}
\end{table}

We first explored the underlying topological characteristics and the significance of KPNs derived from 35 physics examinations, thereby gaining a comprehensive understanding of the characteristics in NCEE physics. The main statistical characteristics of the KPNs are listed in Table~\ref{tab1}. For the reader's convenience, in this article, IKPN represents the network integrating all the KPNs, and $\left\langle \text{KPN} \right\rangle$ represents the result of averaging all the KPNs. In addition, different volumes in various years are abbreviated; for example, 06V1 represents the KPN of National Volume 1 from 2006, where V1 represents the network integrating all the KPNs associated with National Volume 1. In addition, \# of nodes represents the number of nodes in the KPN. Unless otherwise specified, we adopt these notations throughout this article.

The IKPN exhibits substantially larger values across multiple network metrics, including the number of nodes, number of edges, diameter, average degree and average shortest path length, than the individual KPNs. In contrast, the density, transitivity and average clustering coefficient of the IKPN are all lower than those of the other KPNs, indicating its sparser nature. In fact, since the IPKN is an aggregated network consisting of all KPNs, and individual KPN differs due to its emphasis on different knowledge areas, the IKPN knowledge structure is characterized by rich content but weak connectivity. The sparsity of the IPKN, on the other hand, corroborates the density of each individual exam, indicating numerous connections between knowledge points and assessing the learners’ mastery and application of multiple physics knowledge. Furthermore, we examined the KPNs for exams from different volumes and years (Table 1 randomly lists four KPNs, with the remaining results in the Supplementary Table 1), apart from obvious changes in the network's edges, differences in various indicators are relatively minor. These results demonstrate that stable KPNs are obtained regardless of the volume or year. Additionally, statistical analysis of $\left\langle \text{KPN} \right\rangle$ revealed that, on average, a physics examination includes 43 knowledge points and 72 links. Each knowledge point establishes connections with an average of 3.16 other knowledge points, with an average shortest path length of 2, a clustering coefficient of 0.55, a diameter of 4 and a density of 0.075.

Next, we explored several important network effects present in the KPN, including assortative mixing, scale-free characteristics, and the small-world effect.

\begin{figure}
	\centering
	\includegraphics[width=0.9\textwidth]{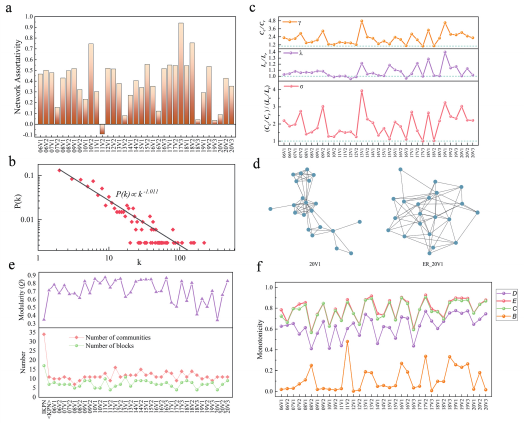}
	\caption{ \textbf{Several important network results of KPNs.}  \textbf{a,} The r-statistics of 35 KPNs. Only 11V2 has a negative R value. \textbf{b,} Degree distribution of IKPN nodes in the integrated network. The degree distribution of the IKPN exhibits a pronounced fat-tail distribution. \textbf{c,} The small-coefficient $\sigma$ is defined as $\sigma = \frac{\left( \frac{C_c}{C_r} \right)}{\left( \frac{L_c}{L_r} \right)}$, where \(C_c\) represents the average clustering coefficient of the largest connected subgraph, \(C_r\) represents the average clustering coefficient of the corresponding ER network, and \(L_r\) is the average shortest path length for the random network. As per the theoretical framework of the small-world effect, if a network's average path length closely resembles that of a random network of equivalent size, and the network's clustering coefficient significantly surpasses that of the corresponding random graph, i.e., $\sigma$ \textgreater 1 ($C_c/C_r \gg 1$ ,and $L_c/L_r \geq 1$) , then the network can be considered a small-world network. \textbf{d,} 20V1 and its corresponding random network graph. It is evident that the network of 20V1 exhibits stronger clustering than its random network. \textbf{e,} Number of communities and modularity values(Q) of the network. Number of blocks represents the number of communities excluding isolated nodes, pairs of nodes and non-closed triangular communities (The three types of communities encompass fewer knowledge points and can roughly represent the number of simple questions in each exam) from the network, reflecting more complex communities. \textbf{f,} Monotonicity values of 4 centrality indexes in 35 Networks. The red curve representing the Eigenvector Centrality $E$ exhibits the highest monotonicity. }
	\label{fig_zh}
\end{figure}

First, the degree correlation offers insight into the ``assortative mixing'' property of a network and is quantified by the simplified Pearson correlation coefficient $r$, with values ranging from [-1, 1] \cite{hu2019segregation, maslov2002specificity}. In our analysis, the $r$ value of the IKPN was 0.0023, indicating weak, albeit positive, assortativity. Further examination of the $r$ values across individual KPNs and visualization via histograms (Fig.~\ref{fig_zh}\textbf{a}) revealed that, except for the 11V2 KPN, all the other KPNs exhibited $r$ values greater than 0. The tendency for nodes with higher degrees in the KPN to connect with highly similar nodes demonstrates the notable assortativity of these networks. This indicates that key knowledge points are consistently associated with each other \cite{catanzaro2004assortative,teller2014emergence,peel2018multiscale}.

Second, to evaluate the degree distribution characteristics of KPNs, we constructed a degree distribution curve for the IKPN. Notably, the degree distribution of the IKPN follows a pronounced fat-tail distribution, characterized by a power law index of $\alpha$ = 1.011, demonstrating scale-free characteristics (Fig.~\ref{fig_zh}\textbf{b}). In scale-free networks, the underlying mechanism governing the adherence of the degree distribution to the power law distribution involves two fundamental principles \cite{barabasi1999emergence}: network growth and the preferential attachment mechanism. As the size of the network increases, new nodes are integrated into the network and form connections with existing nodes, and nodes with higher degrees are more likely to form connections with new nodes. This concept helps explain the evolutionary characteristics of the IKPN. Specifically, new knowledge points are more likely to form connections with higher-degree nodes. Moreover, knowledge points that initially have higher node degrees are more likely to evolve into nodes with even higher degrees in the future. Consequently, classic questions or pivotal knowledge points consistently have more influence over the evolution of examinations.

Then, to investigate whether the KPNs investigated in this article exhibit small-world effects, we established a random model (ER model) with an equivalent number of nodes and edge probabilities for comparative analysis. Network small-worldness has been quantified by a small-coefficient $\sigma$ \cite{humphries2006brainstem,humphries2008network}. Upon careful examination of these 35 KPNs and their corresponding random network attributes, we identified 30 KPNs showing small-world effects (Fig.~\ref{fig_zh}\textbf{c}). Only 5 KPNs did not exhibit small-world characteristics (10V2, 12V1, 12V2, 16V3, 17V3), primarily concentrated in the V2 and V3 types. Randomly selecting KPN 20V1 with a $\sigma$ value greater than 3, we demonstrate its largest connected subgraph alongside its random network graph (Fig.~\ref{fig_zh}\textbf{d}), vividly illustrating the KPN's increased clustering compared to the random network. These findings elucidate the structural complexity and dynamism of knowledge networks.

\begin{figure}
	\centering
	\includegraphics[width=1\textwidth]{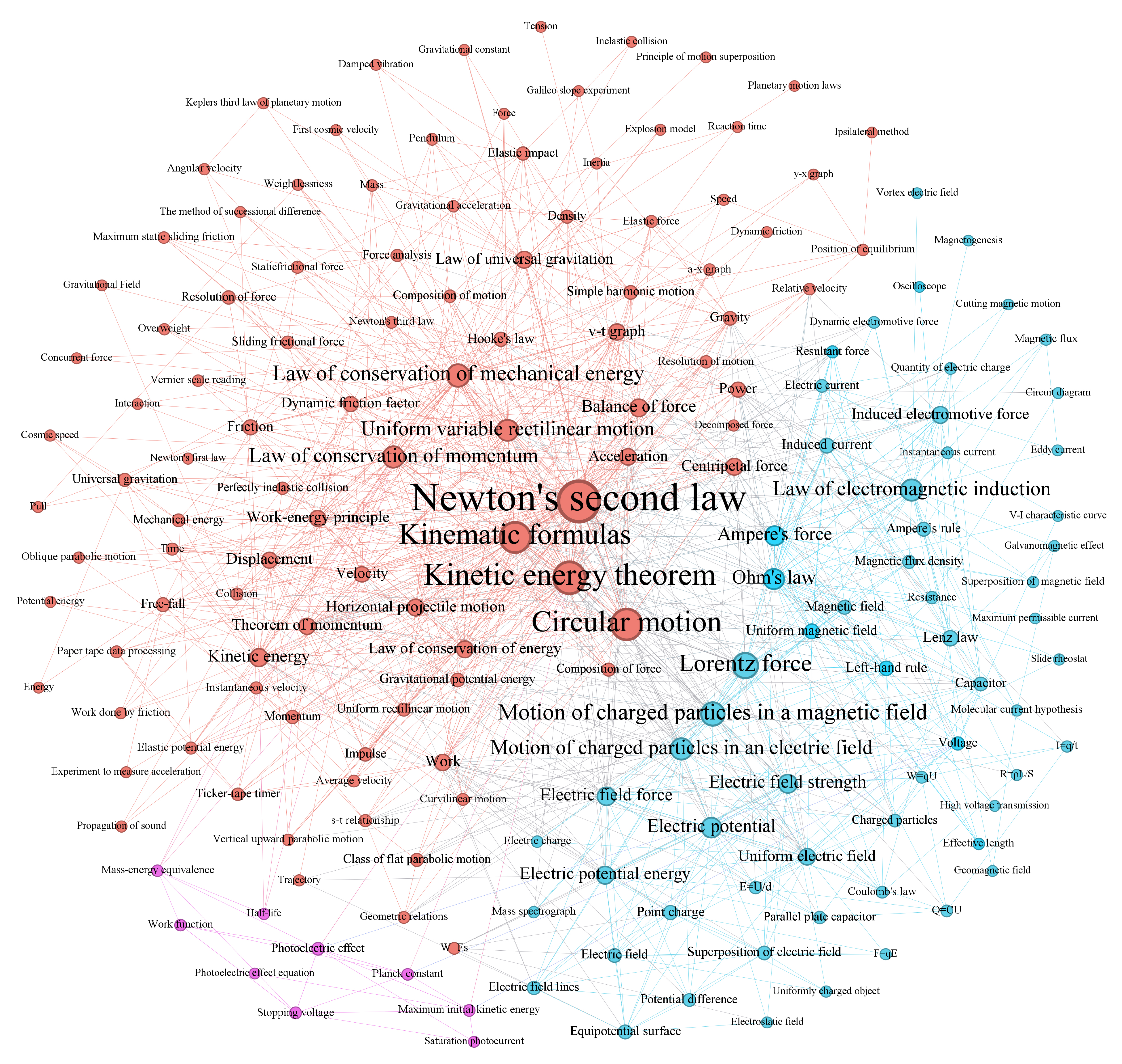}
	\caption{\textbf{The largest community graph representing the IKPN.} The node size is determined according to the degree value. The red nodes represent knowledge related to mechanics, the blue nodes represent knowledge related to electromagnetism, and the purple nodes represent knowledge related to physical optics. The color distinctions in the figure underscore the pronounced clustering features evident within the network.}
	\label{fig2}
\end{figure}

Furthermore, the community structure and key nodes within the KPNs were analyzed. Communities are important structures in complex networks \cite{wasserman1994social, newman2006modularity}. For the KPNs investigated in this study, delineating community structures can elucidate the interplay between knowledge modules within physics examinations and the evolutionary dynamics of physics knowledge modules (knowledge module primarily refers to a set of knowledge points that frequently appear together to address the same question) in college entrance examinations. We utilized the semi-synchronous label propagation algorithm \cite{cordasco2010community} to identify community structures across various KPNs and computed the corresponding modularity values ($Q$) (Fig.~\ref{fig_zh}\textbf{e}). In the IKPN, we reliably identified 34 communities with a network modularity of $Q = 0.35$, with $Q \geq 0.3$ being the empirically determined threshold for accurate representation of real community structures \cite{guimera2005functional}. Among these 34 communities, the largest included 172 members (knowledge points), while the smallest contained only one member. The seven largest communities (Supplementary Table 2) collectively comprised 266 knowledge points, constituting 79.2\% of the IKPN. Together, these communities represent approximately 80\% of high school physics knowledge. Notably, the largest communities depicted in Fig. ~\ref{fig2} correspond to diverse knowledge domains including mechanics, electromagnetism and physical optics. Therefore, the knowledge structure of the IKPN primarily forms seven large knowledge modules. These modules are tightly interconnected internally, with some connections between them, resulting in a clearly discernible structure.

The KPN community detection results revealed little variation in the community division metrics among these KPNs (Fig.~\ref{fig_zh}\textbf{e}). Indeed, we observe similar patterns among the communities identified for the 35 KPNs. The average $Q$ value of the KPNs was 0.7, and a minimum of 0.35 for 20V1, indicating that individual KPN exhibited more pronounced community structures than the IKPN (Supplementary Fig. 1, randomly selecting 12V2 to demonstrate its community structure). Moreover, after removing isolated nodes, pairs of nodes and non-closed triangle communities from the network (the three types of communities encompass fewer knowledge points and can roughly represent the number of simple questions in each exam), statistical analyses based on the community data revealed that the IKPN includes 17 community blocks (community blocks represent closed triangular communities and communities with more than three nodes), while $\left\langle \text{KPN} \right\rangle$ includes approximately 7 community blocks. This reduction by four, compared to the original average of 11 community blocks, suggests that each examination includes approximately four simple questions.

Finally, we also aimed to identify the importance of nodes in the network. Various indices, such as the degree centrality ($D$), eigenvector centrality ($E$), closeness centrality ($C$) and betweenness centrality ($B$), have been proposed to analyse node characteristics from diverse perspectives \cite{aral2012identifying, becker2017network}. The discriminative efficacy of these metrics can be assessed through monotonicity. The IKPN results indicate that eigenvector centrality ($E$) demonstrates the highest monotonicity, with a value of 0.9966. Similar outcomes are observed across the other 35 KPNs (Fig.~\ref{fig_zh}\textbf{f}), where $E$ exhibits the highest monotonicity and thus enables the optimal node differentiation. Consequently, in this study, the $E$ index is leveraged for node sorting within the network to identify pivotal knowledge points. We identified the top 30 knowledge nodes in the IKPN based on eigenvector centrality ($E$) (Supplementary Table 3). These key nodes are predominantly within the domains of mechanics and electromagnetism, with mechanics nodes comprising 60\% of the total. Within mechanics, kinematics and dynamics are considered the most important nodes, while statics is represented only twice among the top 30 nodes. Conversely, electromagnetism comprises 12 nodes, with electricity and magnetism each represented equally, at six times each. The first 30 nodes indicate the pivotal knowledge points within the KPNs and thus represent the most important content within the high school physics curriculum. Similar results were obtained from the analysis of 35 KPNs, with the main nodes still concentrated in mechanics and electromagnetics (The overall color palette focuses on orange and red, Supplementary Fig. 2).

\subsection{Examination difficulty}\label{subsec2}

Measuring the difficulty of examinations is paramount, as the reasonable establishment of exam difficulty directly impacts the effectiveness and fairness of evaluation outcomes. Employing network methods, we propose a novel metric, the comprehensive difficulty coefficient $F_d$, to quantify exam difficulty. This metric encompasses four topological measures: average degree, network density, average clustering coefficient, and network transitivity. This metric, originating solely from the content of examination papers, evaluates exam difficulty based on the intrinsic correlations among knowledge points, considering both breadth and depth of questions, independent of the examination creator and respondent. Prior to formal analysis, we tested the reliability and applicability of the $F_d$ metric (for further details, refer to the Methods). We compared the relationship between the average scores on the physics NCEE over different years in four provinces and changes in $F_d$, finding a direct correlation within each province: the higher the average score is, the lower the comprehensive difficulty coefficient. This confirms the effectiveness of the comprehensive difficulty coefficient $F_d$ in accurately assessing examination difficulty.

Subsequently, we computed difficulty coefficients for a series of KPNs, aiming to reflect exam difficulty from the perspective of networks. Moreover, taking into account the overall evolution and local correlations of examinations, we categorized the 35 examinations based on criteria including the curriculum reform time, national volume number and question type (Supplementary Methods). The corresponding KPNs are were constructed based on these classifications. Based on the proposed difficulty metrics $F_d$, this study explores the characteristics and difficulty features of exam over different periods and types from a long-term, multidimensional perspective.

\subsection{Evolution of KPNs}\label{subsec3}

\begin{table}[h]
	\caption{\textbf{Basic topological quantities of networks in four periods.} First stage represents the network integrating all the KPNs from 2006 to 2010. \# of examinations represents the number of exam papers included in each stage. \# of communities represents the number of communities. $F_d$ represents the comprehensive difficulty coefficient. This metric is multiplied by 10 to output more reasonable numerical values for enhanced readability and interpretability. }	\label{tab_stage}
	\begin{tabular}{@{}lllll@{}}
		\toprule
		Topological quantities         & First stage & Second stage & Third stage & Fourth stage \\
		\midrule
		\# of examinations             & 10          & 8            & 8           & 9            \\
		\# of nodes                    & 168         & 191          & 168         & 187          \\
		\# of edges                    & 428         & 501          & 445         & 664          \\
		\# of communities              & 28          & 35           & 27          & 27           \\
		Diameter                       & 8           & 8            & 7           & 7            \\
		Density                        & 0.031       & 0.028        & 0.032       & 0.038        \\
		Assortativity                  & 0.07        & 0.08         & 0.09        & 0.03         \\
		Transitivity                   & 0.384       & 0.432        & 0.398       & 0.410        \\
		Average degree                 & 5.10        & 5.25         & 5.30        & 7.10         \\
		Average shortest path          & 3.86        & 3.61         & 3.52        & 3.40         \\
		Average clustering coefficient & 0.091       & 0.160        & 0.103       & 0.086        \\
		$F_d$(x10)                     & 0.054       & 0.100        & 0.069       & 0.096        \\
		\bottomrule
	\end{tabular}
\end{table}

Since 2000, Chinese NCEE has undergone four rounds of curriculum reforms, each significantly impacting the examination system. We categorized 35 examinations into four stages based on the release dates of the curriculum reform policies (Supplementary Methods). For each stage, we constructed the corresponding IKPN to explore the developmental trends in NCEE physics throughout these reforms.

\begin{figure}
	\centering
	\includegraphics[width=0.9\textwidth]{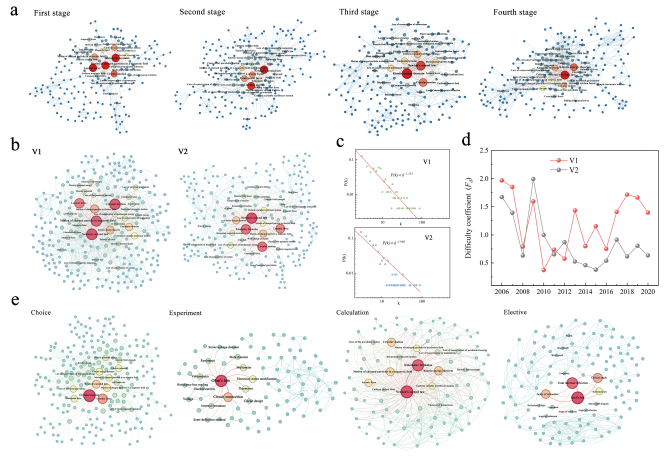}
	\caption{ \textbf{Several results of IKPN across different periods, volumes, and question types.}  \textbf{a,} Network diagrams of the IKPNs for the four stages. Node sizes are determined according to $E$. As the E value increases, the nodes become larger, and their color approaches red. The second stage exhibits strong local clustering effects. The overall network density of the first, third, and fourth stages continues to increase. \textbf{b,} Network diagrams with different volumes. Node sizes are determined according to $E$. V1 is significantly denser than V2, with fewer simple communities and stronger clustering effects. \textbf{c,} The degree distribution graphs for both volumes, V1 KPN and V2 KPN, exhibit scale-free characteristics with a clear fat-tailed distribution. \textbf{d,} Comparative diagrams of the difficulty in two volumes. The difficulty of V1 is consistently higher than V2 for most years, with the exception of the period between 2009 and 2012, during which V1 was less difficult than V2. In terms of quantity, V1 has consistently been proven to be more challenging than V2. Moreover, this characteristic became more pronounced and stable after 2013. \textbf{e,} The network diagrams of the four versions of KPN based on question types. Node sizes are sorted according to the Eigenvector Centrality index $E$. The preferred knowledge points of the four KPNs differ, leading to distinct differences in their network diagrams. Choice question contains the most nodes, while calculation question exhibits the most frequent connections between nodes. }
	\label{fig_2-zh}
\end{figure}

Our analysis of the evolution parameter of KPNs over four stages, corresponding  to the four rounds of curriculum reform undergone by the NCEE since 2000, reveals changes over time in high school physics examination. Table~\ref{tab_stage} indicates notable fluctuations in various network indicators across different stages. Specifically, the average degree progressively increases across the stages, and the average shortest path length decreases. Moreover, the average clustering coefficient and assortativity initially increase and subsequently decrease. Excluding the data from the second stage (which differ considerably), the number of nodes, number of edges, density and transitivity values increase across the remaining three stages. The second stage, despite having only 8 papers, exhibits relatively large or maximum values for node count, edge count, transitivity, etc., with $F_d$ being the highest. These findings suggest that the second stage of reform appears to have considerably increased the difficulty of examinations. Certainly, over the span of 15 years, the overall trend indicates a noticeable increase in the difficulty of high school exams, with the $F_d$ gradually increasing from the first to third and third to fourth stages. Fig.~\ref{fig_2-zh}\textbf{a} also exhibit similar characteristics.

\subsection{Comparison of KPNs for different volumes}\label{subsec4}

\begin{table}[h]
	\caption{\textbf{Basic topological quantities of networks in two volumes.} V1 represents the network integrating all the KPNs associated with National Volume 1. \# of blocks represents the number of communities excluding isolated nodes, pairs of nodes and non-closed triangular communities from the network. }	\label{tab_volume}
	\begin{tabular}{@{}lll@{}}
		\toprule
		Topological quantities         & V1    & V2    \\
		\midrule
		\# of examinations             & 15    & 15    \\
		\#of nodes                     & 252   & 219   \\
		\# of edges                    & 931   & 665   \\
		\# of communities              & 27    & 30    \\
		\# of blocks                   & 18    & 17    \\
		Diameter                       & 7     & 11    \\
		Density                        & 0.029 & 0.028 \\
		Assortativity                  & 0.05  & 0.07  \\
		Transitivity                   & 0.35  & 0.33  \\
		Average degree                 & 7.39  & 6.07  \\
		Average shortest path          & 3.26  & 3.83  \\
		Average clustering coefficient & 0.11  & 0.06  \\
		$F_d$(x10)                     & 0.08  & 0.04  \\
		\bottomrule
	\end{tabular}
\end{table}

As previously mentioned, the examination papers under investigation originate from NCEE. These papers were designed with distinct levels of difficulty. We categorized the 35 papers according to their volumes (Supplementary Methods) and constructed the corresponding IKPN for each volume. This approach allows us to explore whether these different volumes of examination papers exhibit similar characteristics when forming networks.

Table~\ref{tab_volume} and Fig.~\ref{fig_2-zh}\textbf{b} revealed considerable differences in the network structures of IKPNs for different national volumes. Most network indicators, including the number of nodes, number of edges, average degree, network density, network transitivity and average clustering coefficient, are larger for the V1 KPN than for the V2 KPN. The $F_d$ metric proposed in this article shows a similar trend, with a higher value for the V1 KPN. These findings show distinct difficulty profiles for various volumes in physics college entrance examinations, with questions from V1 being more difficult than those from V2. This trend is consistent with the intended design of the NCEE. Moreover, the result reiterates the efficacy of the $F_d$ index in accurately reflecting examination difficulty, thus underscoring the importance of this study.

Furthermore, the network diameter and average shortest path length of the V1 KPN were notably smaller than those of the V2 KPN, suggesting that the V1 KPN has superior network connectivity. Although the V1 KPN has fewer total communities, it includes more blocks, demonstrating that this network contains fewer simplistic questions focused on singular knowledge points. Consequently, these examinations are more difficult, consistent with our earlier observations. In addition, the crucial nodes in the KPNs for both examinations are similar, primarily centered on mechanics and electromagnetism nodes (Supplementary Table 4). There are 12 common nodes, with a similarity rate of 80\%. Notably, the KPNs for both types of examinations exhibit similar assortativity (Fig.~\ref{fig_2-zh}\textbf{c}), displaying scale-free characteristics, with fundamental knowledge points remaining interdependent and retaining stable, prominent positions.

To explore the difficulty differences and changes over time for exams from different volumes, we computed the comprehensive difficulty coefficients for KPNs based on V1 and V2 exams conducted annually from 2006 to 2020 (Fig.~\ref{fig_2-zh}\textbf{d}). Over this 15 years, the V1 exams were more difficult than the V2 exams for 11 of the years; only from 2009 to 2010 and 2012 were the V1 exams less difficulty than the V2 exams. Quantitatively, V1 is consistently more challenging than V2, aligning with earlier research findings. Furthermore, the difference in difficulty between the two types of exams considerably increases after 2013. Notably, during the early stage of this round of basic education curriculum reform, the difficulty of V1 exams was lower than that of V2 exams. Educational reforms inevitably influence question formulation and examination paper composition, which is reflected in the KPN as difficulty ambiguity between volumes. Regarding the change in the exam difficulty, $F_d$ fluctuated between approximately 0.35 and 2. Initially, this metric showed considerable fluctuations, gradually stabilizing in the later stages (the metrics for V1 and V2 stabilizing at approximately 1.5 and 0.75, respectively), suggesting an increasing precision and consistency in the formulation of the exams by examination makers.

\subsection{Comparison of different question types in KPNs}\label{subsec5}

\begin{table}[h]
	\caption{\textbf{Basic topological quantities of networks in four question types.} "Choice" represents  the network integrating all choice questions, also known as the IKPN based on the choice question set. }\label{tab_question}
	\begin{tabular}{@{}lllll@{}}
		\toprule
		Topological quantities         & Choice & Experiment & Calculation & Elective \\
		\midrule
		\# of nodes                    & 228    & 83         & 94          & 125      \\
		\# of edges                    & 881    & 152        & 430         & 307      \\
		\# of communities              & 24     & 18         & 8           & 20       \\
		Diameter                       & 12     & 4          & 5           & 11       \\
		Density                        & 0.03   & 0.04       & 0.10        & 0.04     \\
		Assortativity                  & 0.11   & -0.07      & -0.16       & 0.06     \\
		Transitivity                   & 0.37   & 0.36       & 0.37        & 0.52     \\
		Average degree                 & 7.73   & 3.66       & 9.15        & 4.91     \\
		Average shortest path          & 3.88   & 2.92       & 2.73        & 5.00     \\
		Average clustering coefficient & 0.07   & 0.12       & 0.06        & 0.08     \\
		$F_d$(x10)                     & 0.06   & 0.07       & 0.21        & 0.08     \\
		\bottomrule
	\end{tabular}
\end{table}

The NCEE physics includes choice questions, experimental questions, calculation problems, and elective questions. Considering the inherent knowledge structure characteristics of different question types in the examinations, we categorized the 35 examinations by question type (Supplementary Methods). For each question type, we constructed question-based IKPN to explore the network structure characteristics of different question types.

We next investigated different question types in the constructed knowledge networks (Table~\ref{tab_question} and Fig.~\ref{fig_2-zh}\textbf{e}), observing significant differences in the numbers of nodes and edges across question types (Specific information on question types can be found in the Supplementary Methods and Fig.~\ref{fig_methods}\textbf{a}). The choice questions had the highest assortativity, communities and network diameter but had the lowest network density and comprehensive difficulty coefficient. These results imply that choice questions depend on knowledge point coverage and are thus less difficult.

The experimental questions showed the smallest number of nodes, number of edges, transitivity and average degree but the highest average clustering coefficient. This KPN had a comprehensive difficulty coefficient of 0.07, suggesting a marked directional scope. Although this KPN contains numerous clusters containing knowledge points, the smaller number of knowledge points leads to relatively lower difficulty levels.

In contrast, calculation questions, although comprising only 94 knowledge points (merely 11 more than the experimental question type), showed 430 connecting edges. They exhibit the smallest average shortest path, number of communities and average clustering coefficient but the highest network density, average degree and comprehensive difficulty coefficient. Calculation questions are commonly used to test learners' reasoning and their ability to apply concepts. The maximum average degree increases candidates' potential responses during reasoning and application, consequently increasing question difficulty. Finally, the elective questions demonstrated the highest transitivity and average shortest path. The comprehensive difficulty coefficient of 0.08 is slightly larger than that of both the choice and experimental questions but still considerably lower than that of calculation questions.

Remarkably, we observed differences between the V1 and V2 KPNs for the same question type (Supplementary Table 5). The difficulty of the experimental questions in V1 was similar to that in V2; however, the difficulty of the other question types was notably greater in V1 than in V2. These findings further validate that V1 exams are more difficult than V2 exams.

Using the eigenvector centrality index $E$, we identified the pivotal nodes in KPNs corresponding to various question types (Supplementary Table 6). The important knowledge points of these four networks are considerably different. Notably, the knowledge points of choice and calculation questions considerably similar and mainly involve mechanics and electromagnetism topics. However, their distribution differs. In choice questions, mechanics and electromagnetism are equally represented, whereas calculation questions tend to integrate dynamics from mechanics with electromagnetism more prominently. Therefore, knowledge points related to functional relationships are highlighted to a greater extent. The key nodes for the experimental questions involve electrical experiments, including tasks such as circuit design and voltammetry resistance measurement. Conversely, elective questions mainly involve optical content, which is considerably different than the other question types.

\subsection{The impact of the small-world effect on exam difficulty}\label{subsec5}

Furthermore, we investigated the correlation between the small-world effect and exam difficulty (Supplementary Table 7). We found a positive correlation (0.608, $p < 0.001$) between the comprehensive difficulty coefficient ($F_d$) of the 30 KPNs exhibiting small-world properties and their small-coefficient $\sigma$ value. The more pronounced the small-world effect, the higher the difficulty of the exam. To rule out the influence of different volumes, we separately calculated the correlation between the small-coefficient $\sigma$ and difficulty $F_d$ for each volume of examination (The KPNs of V1 and V2. V3, with only 3 samples displaying small-world properties, are excluded from the discussion). The results were consistent: difficult exams exhibited a more evident small-world effect, indicating that small-world effects increase exam difficulty.

Taken together, our results suggest that the knowledge structure and evolution of the NCEE physics exams can be deeply explored based on networks, enabling comprehensive analysis and quantitative evaluation of examinations over a large-scale, extended time frame, and multiple dimensions.

\section{Discussion}\label{sec3}

Given the current challenges in quantifying examinations, such as subjectivity and scale of assessment, we aimed to comprehensively evaluate examinations using complex networks. By investigating 35 physics examinations from 2006 to 2020, analysing a series of KPNs, and evaluating exam difficulty based on network metrics, this study delves deeply into the structural characteristics and evolution of the NCEE physics, thus providing an objective and comprehensive quantification of examinations through network analysis. Additionally, it demonstrates that KPNs can provide critical insights for educators and are interpretable and valuable for examination analysis and subsequent educational improvements.

The NCEE physics aims to assess students' understanding and application of fundamental physics concepts and laws \cite{etkina2010pedagogical}, with a focus on mechanics and electromagnetism. The current research findings align with this principle. Based on the IKPN eigenvector centrality results, the top 9\% of nodes involve mechanics and electromagnetics topics. The seven largest communities collectively cover approximately 80\% of the knowledge in high school physics curriculums. Due to the evident scale-free nature of the IKPNs, these pivotal knowledge points remain important over time as the exams are refined \cite{barabasi2009scale}. For individual exams, the themes encompassed by each KPN vary slightly, with their core knowledge points consistent with those in the IKPN. High assortativity, observed in multiple systems such as social networks and biological networks \cite{catanzaro2004assortative, teller2014emergence, peel2018multiscale}, is also evident in almost all KPNs in our study. Crucial knowledge points remain interconnected. This analysis can be extended to other subjects and exams. Teachers can leverage these findings in daily instruction to accurately identify core exam themes and key knowledge points, encourages students to focus on these areas, and consider students' understanding and misconceptions regarding these areas to develop effective lesson plans \cite{halim2002science}.

The evolution of KPNs over four distinct periods provides insights for NCEE reform. Over time, the topological structure and difficulty of NCEE physics have continuously changed, with the second phase exhibiting notable distinctiveness. Analyzing the overall trend over the past 15 years, we observe a notable increase in exam difficulty as curriculum reforms have deepened. This indicates that curriculum reforms have encouraged students to increase their knowledge and academic abilities \cite{Jiang2020}, a conclusion consistent with curriculum reform studies in the United States and Hong Kong \cite{moyer2011impact,chan2022eight}. Furthermore, the fluctuation in the exam difficulty gradually stabilizes over time, suggesting that examiners' abilities to tailor exam difficulty appear to be increasingly accurate. However, we also observed that in the early stage of the curriculum reform, the differences among volume types reflected in the networks were not clear (Fig.~\ref{fig_2-zh}\textbf{d}), this seems to be related to the inevitable impact of the reforms on the formulation of exam questions and the compilation of exam papers.

Certainly, there are also findings in revealing the characteristics of different types of exams. First, a study of the IKPNs for different volumes found significant structural differences (Table~\ref{tab_volume}). The values of various network indicators for the V1 KPN are greater than those for the V2 KPN, indicating that V1 is more challenging, consistent with the actual design of the NCEE. This same analysis can be extended to other subjects and used by examination makers for pre-evaluation of exams. Second, different question types are also related to distinct knowledge points, resulting in notable differences in network structures and difficulty levels. When teaching these question types, teachers should adopt corresponding instructional strategies based on the characteristics of each type \cite{rohrer2010recent}. For example, experimental questions should focus on hands-on experiments \cite{ornstein2006frequency}, while calculation questions should emphasize comprehensive exercises. This approach helps students understand the logical and hierarchical relationships between knowledge points in different question types, thereby effectively improving teaching.

Moreover, among the 35 KPNs, a total of 30 networks exhibited notable small-world effects. In examination paper networks with small-world properties, the combination of short paths and high clustering leads to rapid information transmission and relatively dense connections between nodes \cite{latora2001efficient, antiqueira2007strong, li2019scaling}. This implies an increase in the interconnections between knowledge points, thereby increasing the difficulty. Our findings also confirm this (Supplementary Table 7), as exams with small-world attributes present students with greater difficulty and challenge questions to answer. Educators can effectively utilize this feature by reasonably employing exam papers with small-world properties to train and develop students' comprehensive thinking and problem-solving abilities, thereby improving the efficiency and quality of their problem-solving skills \cite{rohrer2010recent, lee2020study}.

The present study nevertheless has numerous limitations. The study was limited by the research methodology. we employed a simple undirected network model in this study. Directed networks could represent directional relationships among knowledge points, making it easier to hierarchically organize knowledge points and reveal structural characteristics such as the hierarchy of the knowledge structure. Additionally, when constructing the KPN, we did not account for the varying weights of different types of questions. The dataset used in this work also presents some limitations. We analysed only the national version of NCEE physics. We lacked local examinations and routine assessments that are independently created questions. Additionally, we focused solely on the subject of physics. To ensure robust results, future research could use datasets from other subjects and different types of examinations, while constructing directed networks and incorporating the weighting of different question types to facilitate a deeper tracing of knowledge structures.

Overall, to our knowledge, this work represents the first attempt to investigate the knowledge structure, statistical regularities, and evolutionary characteristics of the physics NCEE by using a complex network approach. The results show that complex networks could enable comprehensive and meticulous analyses of the knowledge structure and overall characteristics of examinations. In addition, we proposed an indicator based on the intrinsic correlations among knowledge points to assess the difficulty of examinations. This metric is independent of the examination creator and respondent. Through a series of analyses, including preliminary experiments and different exam types, we have demonstrated that this metric effectively reflects exam difficulty. Hence, complex networks should be considered in subsequent educational research involving exam analysis or assessment. Combining the aforementioned findings, using natural language processing (NLP), machine learning \cite{ginev2019scientific}, and deep learning \cite{deng2023chinese, minaee2021deep, kim2020validation} methods can facilitate automatic text processing, knowledge point identification, and examination paper generation \cite{wu2020exam}, enabling in-depth comparison and analysis of exams. Our results offer new perspectives for exam analysis, teaching adjustments and improvements. These insights can be used to enhance exams in physics and other fields, thereby guiding the improvement of teaching practices across various subjects. Our method can be applied to assess exam difficulty in test design, analyze and provide feedback on exams post-assessment. Teachers can leverage the network properties of examination papers to accurately identify core themes and key knowledge, develop corresponding instructional strategies based on different question types, and use exams exhibiting small-world attributes to develop students' abilities. This approach holds promising applications in educational practice.

\section{Methods}\label{sec4}

\subsection{Research materials}

\begin{figure}[h]
	\centering
	\includegraphics[width=0.8\textwidth]{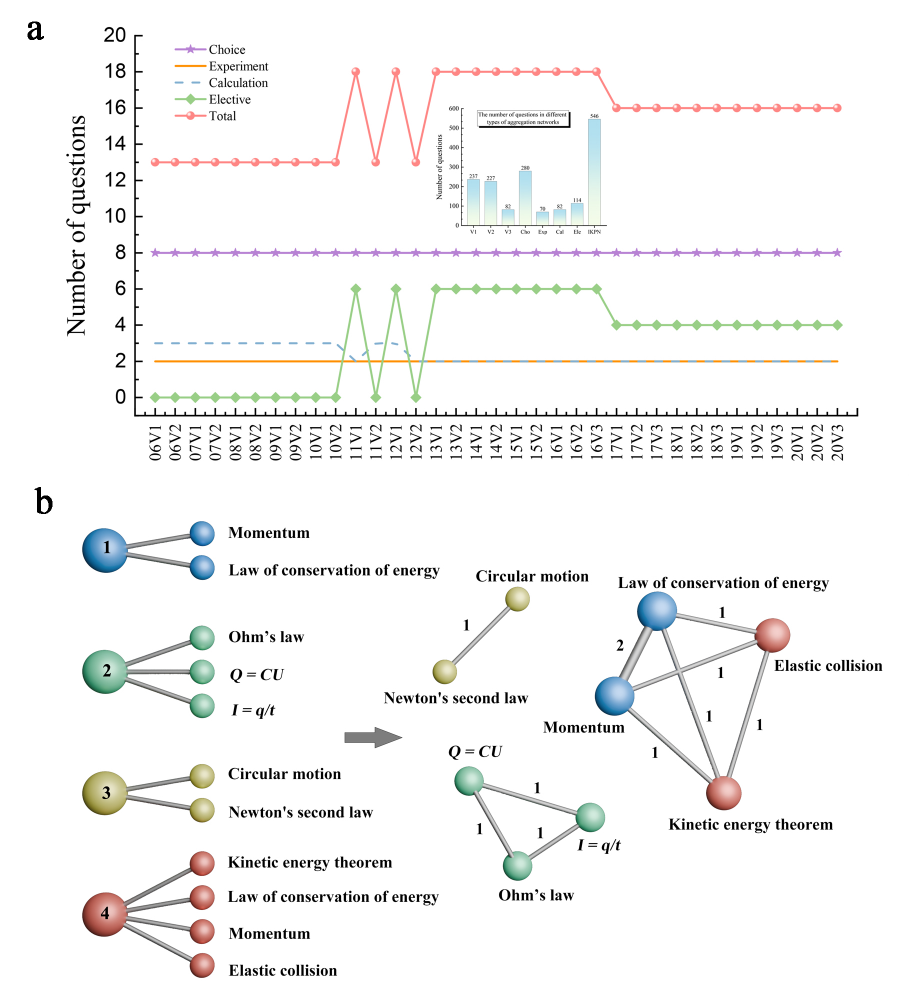}
	\caption{\textbf{Question details and an example of KPN construction.}  \textbf{a,} The number of questions. The line chart displays the number questions in four questions types as well as the total number of questions across 35 examination papers. The central bar chart shows the number of questions in different types of aggregation networks. \textbf{b,} An example of how the knowledge points network (KPN) is constructed. The example network consists of four questions, each marked with blue, green, yellow, and red colors, with each color indicating related knowledge points within the same question. The numbers on the left nodes represent the question numbers, and the numbers on the edges represent the edge weights. Following the principles of network construction, the network graph on the right side was formed. In this graph, the second and third questions are not connected to other questions, forming a single community structure. The first question contains two knowledge points that also appear in the fourth question, establishing a connection between the two questions, resulting in a more complex clustering. In particular, the``Momentum'' and ``Law of conservation of energy'' nodes appear twice within the same question, hence the weight of the edge is 2.}\label{fig_methods}
\end{figure}

To investigate the characteristics and progression of examinations, we selected 35 physics examinations from the Chinese NCEE spanning from 2006 to 2020, focusing on national volumes, as the primary research material. The examination materials were acquired from the relevant authoritative institutions. Based on complex networks, we investigated the structure, types and difficulty distribution of 546 questions across these examinations. The structure and question types, e.g., compulsory questions and elective questions, are consistent across examinations but the number of questions is slightly different. In the examinations from 2006 to 2010, only compulsory questions were included. These compulsory questions included 8 choice questions, 2 laboratory questions and 2-3 comprehensive calculation questions. Elective questions correspond to optional modules in the physics curriculum, prompting candidates to choose and answer one set of questions from a pool of 2-3 options (each option contains two questions). Specific details of the questions are shown in Fig.~\ref{fig_methods}\textbf{a}.

\subsection{Network analysis}
The steps taken in the network analysis described in this article are as follows:

\begin{enumerate}[1.]
	\item[1.] \textbf{Data preprocessing}
\end{enumerate}

Each set of physics examinations was carefully divided into statistical knowledge points. We classified each question according to the question number and listed all knowledge points associated with each question in detail. The knowledge points were defined based on high school physics textbooks published by the People's Education Press and the book `A dictionary of physics' \cite{rennie2019dictionary}. The main content includes the physics concepts, laws and theorems involved in the examination questions. Each node corresponds to a specific knowledge point, and the final data are clearly presented with the volume number, question number and node ID (More information and examples refer to the Supplementary Methods 5). To ensure the reliability of the data, we adopted a rigorous method; that is, two researchers worked together to collectively summarize the knowledge points of the examinations. In the statistical analysis of these knowledge points, if there were any differences, other researchers participated in the discussion until a consensus was reached.

\begin{enumerate}[2.]
	\item[2.] \textbf{Knowledge point network(KPN) construction}
\end{enumerate}

The physics examinations consisted of multiple questions, with each question comprising various knowledge points. Considering knowledge relevance \cite{sizemore2018knowledge}, it can be assumed that knowledge points appearing in the same question are interrelated, with closer relationships between these points than with other points. Therefore, the KPN is constructed according to the following rules: 1) knowledge points appearing in the physics examination are regarded as network nodes; 2) if two knowledge points appear in the same question, an edge exists between them; and 3) network edges are undirected but we consider their weights, and we define the edge weight as the number of times that two knowledge nodes are associated with the same question. Hence, KPNs are undirected weighted networks, as shown in Fig.~\ref{fig_methods}\textbf{b}.

\begin{enumerate}[3.]
	\item[3.] \textbf{Network measures}
\end{enumerate}

Based on the constructed KPN, we systematically investigate the structural properties and evolution of examinations using the topological metrics of complex networks. Complex networks exhibit nontrivial topological features, such as small-world effects \cite{antiqueira2007strong} and scale-free characteristics \cite{barabasi1999emergence}, that are not observed in simple networks \cite{kim2008complex}. A brief introduction to some important network indicators and properties considered in this article is provided in the Supplementary Information (Supplementary Table 8).

\subsection{Difficulty coefficient of exams}

Additionally, assessing examination difficulty through a network science approach is another  purpose of this study. The difficulty of an examination includes two main aspects: breadth and depth \cite{kubinger2007item}. Breadth represents the comprehensiveness of the knowledge points covered in examination questions. In terms of KPNs, a network's breadth can be increased by including more knowledge points. To mitigate the influence of the number of questions on the breadth, the average degree is employed as a metric to reflect the breadth of the KPN. Depth denotes the complexity of questions, emphasizing the connections among knowledge points and the degree of clustering, including both global connectivity and local clustering. Therefore, we can evaluate depth through indicators such as the network density, average clustering coefficient and network transitivity. Considering these factors, we introduce a novel indicator known as the comprehensive difficulty coefficient ($F_d$), with the aim of reflecting examination difficulty using network characteristics. This metric is defined as follows:

\begin{equation} \label{eq:1}
	F_d = \langle k \rangle \times \rho \times T \times \langle  c \rangle
\end{equation}

This indicator comprehensively reflects the breadth and depth of each exam. A larger value implies that there are more nodes, stronger clustering and denser connections in the network, indicating that the questions are related to more knowledge points and thus more complex, indicating a more difficult examination. We also aimed to explore whether the proposed indicator reflects exam difficulty.

Before the formal analysis, a pretest was conducted to assess the reliability and applicability of the comprehensive difficulty coefficient. We obtained the average scores of NCEE physics conducted in four provinces over several years from relevant authoritative organizations, identified as SX, HN, HB and FJ. Utilizing network science theory, we analysed the topological features of the KPNs for these examinations (Supplementary Table 9). Considering the notable impact of different test populations on the NCEE scores across provinces, we compared data from the same province. We observed that within the same province, there was a direct correlation between the average score of the examination and the corresponding comprehensive difficulty coefficient: the lower the average score was, the greater the comprehensive difficulty coefficient. Given that the average score of the examination directly reflects the difficulty of the examination, this finding validates the efficacy of the comprehensive difficulty coefficient in accurately assessing exam difficulty, thus supporting our initial hypothesis. Thus, we employed this indicator to evaluate exam difficulty in our subsequent analyses.

\subsection{Statistical analysis}

Based on the network analysis results obtained above, the average values of various indicators for the 35 Knowledge Point Networks (KPNs) were calculated. Moreover, we define the formula for the overlap rate of key knowledge points as \(\frac{n}{N}\), where \(n\) represents the number of overlapping knowledge points, and \(N\) represents the total number of key knowledge points. The software and packages used in this article include Python 3.10, Gephi and NetworkX.

\section*{Acknowledgements}
This research is supported by the Guangdong Major Project of Basic and Applied Basic Research No. 2020B0301030008, and the National Natural Science Foundation of China (Grant Nos. 62377022, 11935007, 62293552). Thanks to Yu-Ze Tian for the technical support provided in data processing. We also would like to thank Jian-Yao Li, Gao-Jie Li, Yi-Fan Li and Xiao-Yan Li for their help with the data collection.

\section*{Competing interests}
The authors declare no competing interests.

\bibliography{refrence.bib}


\begin{thebibliography}{65}
\ifx \bisbn   \undefined \def \bisbn  #1{ISBN #1}\fi
\ifx \binits  \undefined \def \binits#1{#1}\fi
\ifx \bauthor  \undefined \def \bauthor#1{#1}\fi
\ifx \batitle  \undefined \def \batitle#1{#1}\fi
\ifx \bjtitle  \undefined \def \bjtitle#1{#1}\fi
\ifx \bvolume  \undefined \def \bvolume#1{\textbf{#1}}\fi
\ifx \byear  \undefined \def \byear#1{#1}\fi
\ifx \bissue  \undefined \def \bissue#1{#1}\fi
\ifx \bfpage  \undefined \def \bfpage#1{#1}\fi
\ifx \blpage  \undefined \def \blpage #1{#1}\fi
\ifx \burl  \undefined \def \burl#1{\textsf{#1}}\fi
\ifx \doiurl  \undefined \def \doiurl#1{\url{https://doi.org/#1}}\fi
\ifx \betal  \undefined \def \betal{\textit{et al.}}\fi
\ifx \binstitute  \undefined \def \binstitute#1{#1}\fi
\ifx \binstitutionaled  \undefined \def \binstitutionaled#1{#1}\fi
\ifx \bctitle  \undefined \def \bctitle#1{#1}\fi
\ifx \beditor  \undefined \def \beditor#1{#1}\fi
\ifx \bpublisher  \undefined \def \bpublisher#1{#1}\fi
\ifx \bbtitle  \undefined \def \bbtitle#1{#1}\fi
\ifx \bedition  \undefined \def \bedition#1{#1}\fi
\ifx \bseriesno  \undefined \def \bseriesno#1{#1}\fi
\ifx \blocation  \undefined \def \blocation#1{#1}\fi
\ifx \bsertitle  \undefined \def \bsertitle#1{#1}\fi
\ifx \bsnm \undefined \def \bsnm#1{#1}\fi
\ifx \bsuffix \undefined \def \bsuffix#1{#1}\fi
\ifx \bparticle \undefined \def \bparticle#1{#1}\fi
\ifx \barticle \undefined \def \barticle#1{#1}\fi
\bibcommenthead
\ifx \bconfdate \undefined \def \bconfdate #1{#1}\fi
\ifx \botherref \undefined \def \botherref #1{#1}\fi
\ifx \url \undefined \def \url#1{\textsf{#1}}\fi
\ifx \bchapter \undefined \def \bchapter#1{#1}\fi
\ifx \bbook \undefined \def \bbook#1{#1}\fi
\ifx \bcomment \undefined \def \bcomment#1{#1}\fi
\ifx \oauthor \undefined \def \oauthor#1{#1}\fi
\ifx \citeauthoryear \undefined \def \citeauthoryear#1{#1}\fi
\ifx \endbibitem  \undefined \def \endbibitem {}\fi
\ifx \bconflocation  \undefined \def \bconflocation#1{#1}\fi
\ifx \arxivurl  \undefined \def \arxivurl#1{\textsf{#1}}\fi
\csname PreBibitemsHook\endcsname

\bibitem[\protect\citeauthoryear{Brookhart}{2009}]{brookhart2009assessment}
\begin{bchapter}
\bauthor{\bsnm{Brookhart}, \binits{S.M.}}:
\bctitle{Assessment and examinations}.
In: \bbtitle{International Handbook of Research on Teachers and Teaching},
pp. \bfpage{723}--\blpage{738}.
\bpublisher{Springer},
\blocation{Boston, MA}
(\byear{2009})
\end{bchapter}
\endbibitem

\bibitem[\protect\citeauthoryear{Au}{2007}]{au2007high}
\begin{barticle}
\bauthor{\bsnm{Au}, \binits{W.}}:
\batitle{High-stakes testing and curricular control: A qualitative metasynthesis}.
\bjtitle{Educational researcher}
\bvolume{36}(\bissue{5}),
\bfpage{258}--\blpage{267}
(\byear{2007})
\end{barticle}
\endbibitem

\bibitem[\protect\citeauthoryear{Jacob}{2005}]{jacob2005accountability}
\begin{barticle}
\bauthor{\bsnm{Jacob}, \binits{B.A.}}:
\batitle{Accountability, incentives and behavior: The impact of high-stakes testing in the chicago public schools}.
\bjtitle{Journal of public Economics}
\bvolume{89}(\bissue{5-6}),
\bfpage{761}--\blpage{796}
(\byear{2005})
\end{barticle}
\endbibitem

\bibitem[\protect\citeauthoryear{Von~der Embse et~al.}{2016}]{von2016influence}
\begin{barticle}
\bauthor{\bsnm{Embse}, \binits{N.P.}},
\bauthor{\bsnm{Pendergast}, \binits{L.L.}},
\bauthor{\bsnm{Segool}, \binits{N.}},
\bauthor{\bsnm{Saeki}, \binits{E.}},
\bauthor{\bsnm{Ryan}, \binits{S.}}:
\batitle{The influence of test-based accountability policies on school climate and teacher stress across four states}.
\bjtitle{Teaching and Teacher Education}
\bvolume{59},
\bfpage{492}--\blpage{502}
(\byear{2016})
\end{barticle}
\endbibitem

\bibitem[\protect\citeauthoryear{Fang et~al.}{2022}]{fang2022does}
\begin{barticle}
\bauthor{\bsnm{Fang}, \binits{F.}},
\bauthor{\bsnm{McCall}, \binits{B.}},
\bauthor{\bsnm{Zhong}, \binits{B.}}:
\batitle{How does family background influence students’ choice of subjects for the national college entrance examination?}
\bjtitle{Higher Education Research \& Development}
\bvolume{41}(\bissue{6}),
\bfpage{1885}--\blpage{1899}
(\byear{2022})
\end{barticle}
\endbibitem

\bibitem[\protect\citeauthoryear{McCoubrie}{2004}]{mccoubrie2004improving}
\begin{barticle}
\bauthor{\bsnm{McCoubrie}, \binits{P.}}:
\batitle{Improving the fairness of multiple-choice questions: a literature review}.
\bjtitle{Medical teacher}
\bvolume{26}(\bissue{8}),
\bfpage{709}--\blpage{712}
(\byear{2004})
\end{barticle}
\endbibitem

\bibitem[\protect\citeauthoryear{Mulbar et~al.}{2017}]{mulbar2017analysis}
\begin{botherref}
\oauthor{\bsnm{Mulbar}, \binits{U.}},
\oauthor{\bsnm{Rahman}, \binits{A.}},
\oauthor{\bsnm{Ahmar}, \binits{A.}}:
Analysis of the ability in mathematical problem-solving based on solo taxonomy and cognitive style.
World Transactions on Engineering and Technology Education
\textbf{15}(1)
(2017)
\end{botherref}
\endbibitem

\bibitem[\protect\citeauthoryear{Hallinger et~al.}{2014}]{hallinger2014teacher}
\begin{barticle}
\bauthor{\bsnm{Hallinger}, \binits{P.}},
\bauthor{\bsnm{Heck}, \binits{R.H.}},
\bauthor{\bsnm{Murphy}, \binits{J.}}:
\batitle{Teacher evaluation and school improvement: An analysis of the evidence}.
\bjtitle{Educational Assessment, Evaluation and Accountability}
\bvolume{26},
\bfpage{5}--\blpage{28}
(\byear{2014})
\end{barticle}
\endbibitem

\bibitem[\protect\citeauthoryear{Garg et~al.}{2013}]{garg2013analytical}
\begin{barticle}
\bauthor{\bsnm{Garg}, \binits{R.}},
\bauthor{\bsnm{Saxena}, \binits{D.}},
\bauthor{\bsnm{Shekhawat}, \binits{S.}},
\bauthor{\bsnm{Daga}, \binits{N.}}:
\batitle{Analytical study of written examination papers of undergraduate anatomy: Focus on its content validity}.
\bjtitle{Indian Journal of Basic \& Applied Medical Research}
\bvolume{2}(\bissue{8}),
\bfpage{1110}--\blpage{6}
(\byear{2013})
\end{barticle}
\endbibitem

\bibitem[\protect\citeauthoryear{Han and Xiang}{2024}]{han2024alignment}
\begin{botherref}
\oauthor{\bsnm{Han}, \binits{C.}},
\oauthor{\bsnm{Xiang}, \binits{J.}}:
Alignment analysis between china college entrance examination physics test and curriculum standard based on e-sec model.
International Journal of Science and Mathematics Education,
1--20
(2024)
\end{botherref}
\endbibitem

\bibitem[\protect\citeauthoryear{Akhtar and Saeed}{2020}]{akhtar2020measurement}
\begin{barticle}
\bauthor{\bsnm{Akhtar}, \binits{M.}},
\bauthor{\bsnm{Saeed}, \binits{A.}}:
\batitle{Measurement of essential skills in mathematics: A comparative analysis of ssc (grade-x) and gce (o-level) exam papers.}
\bjtitle{Journal of Education and Educational Development}
\bvolume{7}(\bissue{1}),
\bfpage{103}--\blpage{118}
(\byear{2020})
\end{barticle}
\endbibitem

\bibitem[\protect\citeauthoryear{Rupp et~al.}{2001}]{rupp2001combining}
\begin{barticle}
\bauthor{\bsnm{Rupp}, \binits{A.A.}},
\bauthor{\bsnm{Garcia}, \binits{P.}},
\bauthor{\bsnm{Jamieson}, \binits{J.}}:
\batitle{Combining multiple regression and cart to understand difficulty in second language reading and listening comprehension test items}.
\bjtitle{International Journal of Testing}
\bvolume{1}(\bissue{3-4}),
\bfpage{185}--\blpage{216}
(\byear{2001})
\end{barticle}
\endbibitem

\bibitem[\protect\citeauthoryear{Perkins et~al.}{1995}]{perkins1995predicting}
\begin{barticle}
\bauthor{\bsnm{Perkins}, \binits{K.}},
\bauthor{\bsnm{Gupta}, \binits{L.}},
\bauthor{\bsnm{Tammana}, \binits{R.}}:
\batitle{Predicting item difficulty in a reading comprehension test with an artificial neural network}.
\bjtitle{Language testing}
\bvolume{12}(\bissue{1}),
\bfpage{34}--\blpage{53}
(\byear{1995})
\end{barticle}
\endbibitem

\bibitem[\protect\citeauthoryear{Bi et~al.}{2024}]{bi2024difficulty}
\begin{barticle}
\bauthor{\bsnm{Bi}, \binits{S.}},
\bauthor{\bsnm{Liu}, \binits{J.}},
\bauthor{\bsnm{Miao}, \binits{Z.}},
\bauthor{\bsnm{Min}, \binits{Q.}}:
\batitle{Difficulty-controllable question generation over knowledge graphs: A counterfactual reasoning approach}.
\bjtitle{Information Processing \& Management}
\bvolume{61}(\bissue{4}),
\bfpage{103721}
(\byear{2024})
\end{barticle}
\endbibitem

\bibitem[\protect\citeauthoryear{Jiang et~al.}{2014}]{jiang2014structure}
\begin{barticle}
\bauthor{\bsnm{Jiang}, \binits{X.}},
\bauthor{\bsnm{Chen}, \binits{T.}},
\bauthor{\bsnm{Zheng}, \binits{B.}}:
\batitle{Structure of local interactions in complex financial dynamics}.
\bjtitle{Scientific reports}
\bvolume{4}(\bissue{1}),
\bfpage{5321}
(\byear{2014})
\end{barticle}
\endbibitem

\bibitem[\protect\citeauthoryear{Sun et~al.}{2020}]{sun2020revealing}
\begin{barticle}
\bauthor{\bsnm{Sun}, \binits{J.}},
\bauthor{\bsnm{Feng}, \binits{L.}},
\bauthor{\bsnm{Xie}, \binits{J.}},
\bauthor{\bsnm{Ma}, \binits{X.}},
\bauthor{\bsnm{Wang}, \binits{D.}},
\bauthor{\bsnm{Hu}, \binits{Y.}}:
\batitle{Revealing the predictability of intrinsic structure in complex networks}.
\bjtitle{Nature communications}
\bvolume{11}(\bissue{1}),
\bfpage{574}
(\byear{2020})
\end{barticle}
\endbibitem

\bibitem[\protect\citeauthoryear{Fox et~al.}{2020}]{fox2020intrinsic}
\begin{barticle}
\bauthor{\bsnm{Fox}, \binits{K.C.}},
\bauthor{\bsnm{Shi}, \binits{L.}},
\bauthor{\bsnm{Baek}, \binits{S.}},
\bauthor{\bsnm{Raccah}, \binits{O.}},
\bauthor{\bsnm{Foster}, \binits{B.L.}},
\bauthor{\bsnm{Saha}, \binits{S.}},
\bauthor{\bsnm{Margulies}, \binits{D.S.}},
\bauthor{\bsnm{Kucyi}, \binits{A.}},
\bauthor{\bsnm{Parvizi}, \binits{J.}}:
\batitle{Intrinsic network architecture predicts the effects elicited by intracranial electrical stimulation of the human brain}.
\bjtitle{Nature human behaviour}
\bvolume{4}(\bissue{10}),
\bfpage{1039}--\blpage{1052}
(\byear{2020})
\end{barticle}
\endbibitem

\bibitem[\protect\citeauthoryear{Forsman et~al.}{2014}]{forsman2014extending}
\begin{barticle}
\bauthor{\bsnm{Forsman}, \binits{J.}},
\bauthor{\bsnm{Moll}, \binits{R.}},
\bauthor{\bsnm{Linder}, \binits{C.}}:
\batitle{Extending the theoretical framing for physics education research: An illustrative application of complexity science}.
\bjtitle{Physical Review Special Topics-Physics Education Research}
\bvolume{10}(\bissue{2}),
\bfpage{020122}
(\byear{2014})
\end{barticle}
\endbibitem

\bibitem[\protect\citeauthoryear{Lu et~al.}{2020}]{lu2020diversities}
\begin{barticle}
\bauthor{\bsnm{Lu}, \binits{X.}},
\bauthor{\bsnm{Liu}, \binits{X.W.}},
\bauthor{\bsnm{Zhang}, \binits{W.}}:
\batitle{Diversities of learners' interactions in different mooc courses: How these diversities affects communication in learning}.
\bjtitle{Computers \& Education}
\bvolume{151},
\bfpage{103873}
(\byear{2020})
\end{barticle}
\endbibitem

\bibitem[\protect\citeauthoryear{Wolf et~al.}{2021}]{wolf2021complex}
\begin{barticle}
\bauthor{\bsnm{Wolf}, \binits{S.}},
\bauthor{\bsnm{Gonzalez~Canche}, \binits{M.S.}},
\bauthor{\bsnm{Coe}, \binits{K.}}:
\batitle{A complex systems network approach to quantifying peer effects: Evidence from ghanaian preprimary classrooms}.
\bjtitle{Child Development}
\bvolume{92}(\bissue{6}),
\bfpage{1242}--\blpage{1259}
(\byear{2021})
\end{barticle}
\endbibitem

\bibitem[\protect\citeauthoryear{Goh et~al.}{2014}]{goh2014complex}
\begin{barticle}
\bauthor{\bsnm{Goh}, \binits{W.P.}},
\bauthor{\bsnm{Kwek}, \binits{D.}},
\bauthor{\bsnm{Hogan}, \binits{D.}},
\bauthor{\bsnm{Cheong}, \binits{S.A.}}:
\batitle{Complex network analysis of teaching practices}.
\bjtitle{EPJ Data Science}
\bvolume{3},
\bfpage{1}--\blpage{16}
(\byear{2014})
\end{barticle}
\endbibitem

\bibitem[\protect\citeauthoryear{Shu and Gu}{2018}]{shu2018determining}
\begin{barticle}
\bauthor{\bsnm{Shu}, \binits{H.}},
\bauthor{\bsnm{Gu}, \binits{X.}}:
\batitle{Determining the differences between online and face-to-face student--group interactions in a blended learning course}.
\bjtitle{The Internet and Higher Education}
\bvolume{39},
\bfpage{13}--\blpage{21}
(\byear{2018})
\end{barticle}
\endbibitem

\bibitem[\protect\citeauthoryear{Feng and Kirkley}{2020}]{feng2020mixing}
\begin{barticle}
\bauthor{\bsnm{Feng}, \binits{S.}},
\bauthor{\bsnm{Kirkley}, \binits{A.}}:
\batitle{Mixing patterns in interdisciplinary co-authorship networks at multiple scales}.
\bjtitle{Scientific Reports}
\bvolume{10}(\bissue{1}),
\bfpage{7731}
(\byear{2020})
\end{barticle}
\endbibitem

\bibitem[\protect\citeauthoryear{Ramirez-Arellano}{2019}]{ramirez2019students}
\begin{barticle}
\bauthor{\bsnm{Ramirez-Arellano}, \binits{A.}}:
\batitle{Students learning pathways in higher blended education: An analysis of complex networks perspective}.
\bjtitle{Computers \& Education}
\bvolume{141},
\bfpage{103634}
(\byear{2019})
\end{barticle}
\endbibitem

\bibitem[\protect\citeauthoryear{Yun and Park}{2018}]{yun2018extraction}
\begin{barticle}
\bauthor{\bsnm{Yun}, \binits{E.}},
\bauthor{\bsnm{Park}, \binits{Y.}}:
\batitle{Extraction of scientific semantic networks from science textbooks and comparison with science teachers’ spoken language by text network analysis}.
\bjtitle{International Journal of Science Education}
\bvolume{40}(\bissue{17}),
\bfpage{2118}--\blpage{2136}
(\byear{2018})
\end{barticle}
\endbibitem

\bibitem[\protect\citeauthoryear{Stella et~al.}{2017}]{stella2017multiplex}
\begin{barticle}
\bauthor{\bsnm{Stella}, \binits{M.}},
\bauthor{\bsnm{Beckage}, \binits{N.M.}},
\bauthor{\bsnm{Brede}, \binits{M.}}:
\batitle{Multiplex lexical networks reveal patterns in early word acquisition in children}.
\bjtitle{Scientific reports}
\bvolume{7}(\bissue{1}),
\bfpage{46730}
(\byear{2017})
\end{barticle}
\endbibitem

\bibitem[\protect\citeauthoryear{Sizemore et~al.}{2018}]{sizemore2018knowledge}
\begin{barticle}
\bauthor{\bsnm{Sizemore}, \binits{A.E.}},
\bauthor{\bsnm{Karuza}, \binits{E.A.}},
\bauthor{\bsnm{Giusti}, \binits{C.}},
\bauthor{\bsnm{Bassett}, \binits{D.S.}}:
\batitle{Knowledge gaps in the early growth of semantic feature networks}.
\bjtitle{Nature human behaviour}
\bvolume{2}(\bissue{9}),
\bfpage{682}--\blpage{692}
(\byear{2018})
\end{barticle}
\endbibitem

\bibitem[\protect\citeauthoryear{Zhang}{2016}]{zhang2016national}
\begin{bbook}
\bauthor{\bsnm{Zhang}, \binits{Y.}}:
\bbtitle{National College Entrance Exam in China: Perspectives on Education Quality and Equity}.
\bpublisher{Springer},
\blocation{Singapore}
(\byear{2016})
\end{bbook}
\endbibitem

\bibitem[\protect\citeauthoryear{Hu et~al.}{2019}]{hu2019segregation}
\begin{barticle}
\bauthor{\bsnm{Hu}, \binits{J.}},
\bauthor{\bsnm{Zhang}, \binits{Q.-M.}},
\bauthor{\bsnm{Zhou}, \binits{T.}}:
\batitle{Segregation in religion networks}.
\bjtitle{EPJ Data Science}
\bvolume{8}(\bissue{1}),
\bfpage{6}
(\byear{2019})
\end{barticle}
\endbibitem

\bibitem[\protect\citeauthoryear{Maslov and Sneppen}{2002}]{maslov2002specificity}
\begin{barticle}
\bauthor{\bsnm{Maslov}, \binits{S.}},
\bauthor{\bsnm{Sneppen}, \binits{K.}}:
\batitle{Specificity and stability in topology of protein networks}.
\bjtitle{Science}
\bvolume{296}(\bissue{5569}),
\bfpage{910}--\blpage{913}
(\byear{2002})
\end{barticle}
\endbibitem

\bibitem[\protect\citeauthoryear{Catanzaro et~al.}{2004}]{catanzaro2004assortative}
\begin{barticle}
\bauthor{\bsnm{Catanzaro}, \binits{M.}},
\bauthor{\bsnm{Caldarelli}, \binits{G.}},
\bauthor{\bsnm{Pietronero}, \binits{L.}}:
\batitle{Assortative model for social networks}.
\bjtitle{Physical review e}
\bvolume{70}(\bissue{3}),
\bfpage{037101}
(\byear{2004})
\end{barticle}
\endbibitem

\bibitem[\protect\citeauthoryear{Teller et~al.}{2014}]{teller2014emergence}
\begin{barticle}
\bauthor{\bsnm{Teller}, \binits{S.}},
\bauthor{\bsnm{Granell}, \binits{C.}},
\bauthor{\bsnm{De~Domenico}, \binits{M.}},
\bauthor{\bsnm{Soriano}, \binits{J.}},
\bauthor{\bsnm{Gomez}, \binits{S.}},
\bauthor{\bsnm{Arenas}, \binits{A.}}:
\batitle{Emergence of assortative mixing between clusters of cultured neurons}.
\bjtitle{PLoS computational biology}
\bvolume{10}(\bissue{9}),
\bfpage{1003796}
(\byear{2014})
\end{barticle}
\endbibitem

\bibitem[\protect\citeauthoryear{Peel et~al.}{2018}]{peel2018multiscale}
\begin{barticle}
\bauthor{\bsnm{Peel}, \binits{L.}},
\bauthor{\bsnm{Delvenne}, \binits{J.-C.}},
\bauthor{\bsnm{Lambiotte}, \binits{R.}}:
\batitle{Multiscale mixing patterns in networks}.
\bjtitle{Proceedings of the National Academy of Sciences}
\bvolume{115}(\bissue{16}),
\bfpage{4057}--\blpage{4062}
(\byear{2018})
\end{barticle}
\endbibitem

\bibitem[\protect\citeauthoryear{Barab{\'a}si and Albert}{1999}]{barabasi1999emergence}
\begin{barticle}
\bauthor{\bsnm{Barab{\'a}si}, \binits{A.-L.}},
\bauthor{\bsnm{Albert}, \binits{R.}}:
\batitle{Emergence of scaling in random networks}.
\bjtitle{science}
\bvolume{286}(\bissue{5439}),
\bfpage{509}--\blpage{512}
(\byear{1999})
\end{barticle}
\endbibitem

\bibitem[\protect\citeauthoryear{Humphries et~al.}{2006}]{humphries2006brainstem}
\begin{barticle}
\bauthor{\bsnm{Humphries}, \binits{M.D.}},
\bauthor{\bsnm{Gurney}, \binits{K.}},
\bauthor{\bsnm{Prescott}, \binits{T.J.}}:
\batitle{The brainstem reticular formation is a small-world, not scale-free, network}.
\bjtitle{Proceedings of the Royal Society B: Biological Sciences}
\bvolume{273}(\bissue{1585}),
\bfpage{503}--\blpage{511}
(\byear{2006})
\end{barticle}
\endbibitem

\bibitem[\protect\citeauthoryear{Humphries and Gurney}{2008}]{humphries2008network}
\begin{barticle}
\bauthor{\bsnm{Humphries}, \binits{M.D.}},
\bauthor{\bsnm{Gurney}, \binits{K.}}:
\batitle{Network ‘small-world-ness’: a quantitative method for determining canonical network equivalence}.
\bjtitle{PloS one}
\bvolume{3}(\bissue{4}),
\bfpage{0002051}
(\byear{2008})
\end{barticle}
\endbibitem

\bibitem[\protect\citeauthoryear{Wasserman and Faust}{1994}]{wasserman1994social}
\begin{botherref}
\oauthor{\bsnm{Wasserman}, \binits{S.}},
\oauthor{\bsnm{Faust}, \binits{K.}}:
Social network analysis: Methods and applications
(1994)
\end{botherref}
\endbibitem

\bibitem[\protect\citeauthoryear{Newman}{2006}]{newman2006modularity}
\begin{barticle}
\bauthor{\bsnm{Newman}, \binits{M.E.}}:
\batitle{Modularity and community structure in networks}.
\bjtitle{Proceedings of the national academy of sciences}
\bvolume{103}(\bissue{23}),
\bfpage{8577}--\blpage{8582}
(\byear{2006})
\end{barticle}
\endbibitem

\bibitem[\protect\citeauthoryear{Cordasco and Gargano}{2010}]{cordasco2010community}
\begin{bchapter}
\bauthor{\bsnm{Cordasco}, \binits{G.}},
\bauthor{\bsnm{Gargano}, \binits{L.}}:
\bctitle{Community detection via semi-synchronous label propagation algorithms}.
In: \bbtitle{2010 IEEE International Workshop On: Business Applications of Social Network Analysis (BASNA)},
pp. \bfpage{1}--\blpage{8}
(\byear{2010}).
\bcomment{IEEE}
\end{bchapter}
\endbibitem

\bibitem[\protect\citeauthoryear{Guimera and Nunes~Amaral}{2005}]{guimera2005functional}
\begin{barticle}
\bauthor{\bsnm{Guimera}, \binits{R.}},
\bauthor{\bsnm{Nunes~Amaral}, \binits{L.A.}}:
\batitle{Functional cartography of complex metabolic networks}.
\bjtitle{nature}
\bvolume{433}(\bissue{7028}),
\bfpage{895}--\blpage{900}
(\byear{2005})
\end{barticle}
\endbibitem

\bibitem[\protect\citeauthoryear{Aral and Walker}{2012}]{aral2012identifying}
\begin{barticle}
\bauthor{\bsnm{Aral}, \binits{S.}},
\bauthor{\bsnm{Walker}, \binits{D.}}:
\batitle{Identifying influential and susceptible members of social networks}.
\bjtitle{Science}
\bvolume{337}(\bissue{6092}),
\bfpage{337}--\blpage{341}
(\byear{2012})
\end{barticle}
\endbibitem

\bibitem[\protect\citeauthoryear{Becker et~al.}{2017}]{becker2017network}
\begin{barticle}
\bauthor{\bsnm{Becker}, \binits{J.}},
\bauthor{\bsnm{Brackbill}, \binits{D.}},
\bauthor{\bsnm{Centola}, \binits{D.}}:
\batitle{Network dynamics of social influence in the wisdom of crowds}.
\bjtitle{Proceedings of the national academy of sciences}
\bvolume{114}(\bissue{26}),
\bfpage{5070}--\blpage{5076}
(\byear{2017})
\end{barticle}
\endbibitem

\bibitem[\protect\citeauthoryear{Etkina}{2010}]{etkina2010pedagogical}
\begin{barticle}
\bauthor{\bsnm{Etkina}, \binits{E.}}:
\batitle{Pedagogical content knowledge and preparation of high school physics teachers}.
\bjtitle{Physical Review Special Topics-Physics Education Research}
\bvolume{6}(\bissue{2}),
\bfpage{020110}
(\byear{2010})
\end{barticle}
\endbibitem

\bibitem[\protect\citeauthoryear{Barab{\'a}si}{2009}]{barabasi2009scale}
\begin{barticle}
\bauthor{\bsnm{Barab{\'a}si}, \binits{A.-L.}}:
\batitle{Scale-free networks: a decade and beyond}.
\bjtitle{science}
\bvolume{325}(\bissue{5939}),
\bfpage{412}--\blpage{413}
(\byear{2009})
\end{barticle}
\endbibitem

\bibitem[\protect\citeauthoryear{Halim and Meerah}{2002}]{halim2002science}
\begin{barticle}
\bauthor{\bsnm{Halim}, \binits{L.}},
\bauthor{\bsnm{Meerah}, \binits{S.M.M.}}:
\batitle{Science trainee teachers' pedagogical content knowledge and its influence on physics teaching}.
\bjtitle{Research in Science \& Technological Education}
\bvolume{20}(\bissue{2}),
\bfpage{215}--\blpage{225}
(\byear{2002})
\end{barticle}
\endbibitem

\bibitem[\protect\citeauthoryear{Jiang and Guo}{2020}]{Jiang2020}
\begin{bchapter}
\bauthor{\bsnm{Jiang}, \binits{Q.}},
\bauthor{\bsnm{Guo}, \binits{X.}}:
\bctitle{Research on the reform of chinese college entrance examination system}.
In: \bbtitle{Third International Conference on Social Science, Public Health and Education (SSPHE 2019)},
pp. \bfpage{107}--\blpage{111}
(\byear{2020}).
\bcomment{Atlantis Press}
\end{bchapter}
\endbibitem

\bibitem[\protect\citeauthoryear{Moyer et~al.}{2011}]{moyer2011impact}
\begin{barticle}
\bauthor{\bsnm{Moyer}, \binits{J.C.}},
\bauthor{\bsnm{Cai}, \binits{J.}},
\bauthor{\bsnm{Wang}, \binits{N.}},
\bauthor{\bsnm{Nie}, \binits{B.}}:
\batitle{Impact of curriculum reform: Evidence of change in classroom practice in the united states}.
\bjtitle{International Journal of Educational Research}
\bvolume{50}(\bissue{2}),
\bfpage{87}--\blpage{99}
(\byear{2011})
\end{barticle}
\endbibitem

\bibitem[\protect\citeauthoryear{Chan and Luk}{2022}]{chan2022eight}
\begin{barticle}
\bauthor{\bsnm{Chan}, \binits{C.K.}},
\bauthor{\bsnm{Luk}, \binits{L.Y.}}:
\batitle{Eight years after the 3-3-4 curriculum reform: The current state of undergraduates’ holistic competency development in hong kong}.
\bjtitle{Studies in Educational Evaluation}
\bvolume{74},
\bfpage{101168}
(\byear{2022})
\end{barticle}
\endbibitem

\bibitem[\protect\citeauthoryear{Rohrer and Pashler}{2010}]{rohrer2010recent}
\begin{barticle}
\bauthor{\bsnm{Rohrer}, \binits{D.}},
\bauthor{\bsnm{Pashler}, \binits{H.}}:
\batitle{Recent research on human learning challenges conventional instructional strategies}.
\bjtitle{Educational Researcher}
\bvolume{39}(\bissue{5}),
\bfpage{406}--\blpage{412}
(\byear{2010})
\end{barticle}
\endbibitem

\bibitem[\protect\citeauthoryear{Ornstein}{2006}]{ornstein2006frequency}
\begin{barticle}
\bauthor{\bsnm{Ornstein}, \binits{A.}}:
\batitle{The frequency of hands-on experimentation and student attitudes toward science: A statistically significant relation (2005-51-ornstein)}.
\bjtitle{Journal of Science Education and Technology}
\bvolume{15},
\bfpage{285}--\blpage{297}
(\byear{2006})
\end{barticle}
\endbibitem

\bibitem[\protect\citeauthoryear{Latora and Marchiori}{2001}]{latora2001efficient}
\begin{barticle}
\bauthor{\bsnm{Latora}, \binits{V.}},
\bauthor{\bsnm{Marchiori}, \binits{M.}}:
\batitle{Efficient behavior of small-world networks}.
\bjtitle{Physical review letters}
\bvolume{87}(\bissue{19}),
\bfpage{198701}
(\byear{2001})
\end{barticle}
\endbibitem

\bibitem[\protect\citeauthoryear{Antiqueira et~al.}{2007}]{antiqueira2007strong}
\begin{barticle}
\bauthor{\bsnm{Antiqueira}, \binits{L.}},
\bauthor{\bsnm{Nunes}, \binits{M.d.G.V.}},
\bauthor{\bsnm{Oliveira~Jr}, \binits{O.}},
\bauthor{\bsnm{Costa}, \binits{L.d.F.}}:
\batitle{Strong correlations between text quality and complex networks features}.
\bjtitle{Physica A: Statistical Mechanics and its Applications}
\bvolume{373},
\bfpage{811}--\blpage{820}
(\byear{2007})
\end{barticle}
\endbibitem

\bibitem[\protect\citeauthoryear{Li et~al.}{2019}]{li2019scaling}
\begin{bchapter}
\bauthor{\bsnm{Li}, \binits{W.}},
\bauthor{\bsnm{Qiao}, \binits{M.}},
\bauthor{\bsnm{Qin}, \binits{L.}},
\bauthor{\bsnm{Zhang}, \binits{Y.}},
\bauthor{\bsnm{Chang}, \binits{L.}},
\bauthor{\bsnm{Lin}, \binits{X.}}:
\bctitle{Scaling distance labeling on small-world networks}.
In: \bbtitle{Proceedings of the 2019 International Conference on Management of Data},
pp. \bfpage{1060}--\blpage{1077}
(\byear{2019})
\end{bchapter}
\endbibitem

\bibitem[\protect\citeauthoryear{Lee and Lee}{2020}]{lee2020study}
\begin{barticle}
\bauthor{\bsnm{Lee}, \binits{B.}},
\bauthor{\bsnm{Lee}, \binits{Y.}}:
\batitle{A study examining the effects of a training program focused on problem-solving skills for young adults}.
\bjtitle{Thinking Skills and Creativity}
\bvolume{37},
\bfpage{100692}
(\byear{2020})
\end{barticle}
\endbibitem

\bibitem[\protect\citeauthoryear{Ginev and Miller}{2019}]{ginev2019scientific}
\begin{botherref}
\oauthor{\bsnm{Ginev}, \binits{D.}},
\oauthor{\bsnm{Miller}, \binits{B.R.}}:
Scientific statement classification over arxiv. org.
arXiv preprint arXiv:1908.10993
(2019)
\end{botherref}
\endbibitem

\bibitem[\protect\citeauthoryear{Deng et~al.}{2023}]{deng2023chinese}
\begin{botherref}
\oauthor{\bsnm{Deng}, \binits{S.}},
\oauthor{\bsnm{Li}, \binits{Q.}},
\oauthor{\bsnm{Dai}, \binits{R.}},
\oauthor{\bsnm{Wei}, \binits{S.}},
\oauthor{\bsnm{Wu}, \binits{D.}},
\oauthor{\bsnm{He}, \binits{Y.}},
\oauthor{\bsnm{Wu}, \binits{X.}}:
A chinese power text classification algorithm based on deep active learning.
Applied Soft Computing,
111067
(2023)
\end{botherref}
\endbibitem

\bibitem[\protect\citeauthoryear{Minaee et~al.}{2021}]{minaee2021deep}
\begin{barticle}
\bauthor{\bsnm{Minaee}, \binits{S.}},
\bauthor{\bsnm{Kalchbrenner}, \binits{N.}},
\bauthor{\bsnm{Cambria}, \binits{E.}},
\bauthor{\bsnm{Nikzad}, \binits{N.}},
\bauthor{\bsnm{Chenaghlu}, \binits{M.}},
\bauthor{\bsnm{Gao}, \binits{J.}}:
\batitle{Deep learning--based text classification: a comprehensive review}.
\bjtitle{ACM computing surveys (CSUR)}
\bvolume{54}(\bissue{3}),
\bfpage{1}--\blpage{40}
(\byear{2021})
\end{barticle}
\endbibitem

\bibitem[\protect\citeauthoryear{Kim et~al.}{2020}]{kim2020validation}
\begin{barticle}
\bauthor{\bsnm{Kim}, \binits{Y.}},
\bauthor{\bsnm{Lee}, \binits{J.H.}},
\bauthor{\bsnm{Choi}, \binits{S.}},
\bauthor{\bsnm{Lee}, \binits{J.M.}},
\bauthor{\bsnm{Kim}, \binits{J.-H.}},
\bauthor{\bsnm{Seok}, \binits{J.}},
\bauthor{\bsnm{Joo}, \binits{H.J.}}:
\batitle{Validation of deep learning natural language processing algorithm for keyword extraction from pathology reports in electronic health records}.
\bjtitle{Scientific reports}
\bvolume{10}(\bissue{1}),
\bfpage{20265}
(\byear{2020})
\end{barticle}
\endbibitem

\bibitem[\protect\citeauthoryear{Wu et~al.}{2020}]{wu2020exam}
\begin{barticle}
\bauthor{\bsnm{Wu}, \binits{Z.}},
\bauthor{\bsnm{He}, \binits{T.}},
\bauthor{\bsnm{Mao}, \binits{C.}},
\bauthor{\bsnm{Huang}, \binits{C.}}:
\batitle{Exam paper generation based on performance prediction of student group}.
\bjtitle{Information Sciences}
\bvolume{532},
\bfpage{72}--\blpage{90}
(\byear{2020})
\end{barticle}
\endbibitem

\bibitem[\protect\citeauthoryear{Rennie and Law}{2019}]{rennie2019dictionary}
\begin{bbook}
\bauthor{\bsnm{Rennie}, \binits{R.}},
\bauthor{\bsnm{Law}, \binits{J.}}:
\bbtitle{A Dictionary of Physics}.
\bpublisher{Oxford University Press},
\blocation{London}
(\byear{2019})
\end{bbook}
\endbibitem

\bibitem[\protect\citeauthoryear{Kim and Wilhelm}{2008}]{kim2008complex}
\begin{barticle}
\bauthor{\bsnm{Kim}, \binits{J.}},
\bauthor{\bsnm{Wilhelm}, \binits{T.}}:
\batitle{What is a complex graph?}
\bjtitle{Physica A: Statistical Mechanics and its Applications}
\bvolume{387}(\bissue{11}),
\bfpage{2637}--\blpage{2652}
(\byear{2008})
\end{barticle}
\endbibitem

\bibitem[\protect\citeauthoryear{Kubinger and Gottschall}{2007}]{kubinger2007item}
\begin{barticle}
\bauthor{\bsnm{Kubinger}, \binits{K.D.}},
\bauthor{\bsnm{Gottschall}, \binits{C.H.}}:
\batitle{Item difficulty of multiple choice tests dependant on different item response formats--an experiment in fundamental research on psychological assessment}.
\bjtitle{Psychology science}
\bvolume{49}(\bissue{4}),
\bfpage{361}
(\byear{2007})
\end{barticle}
\endbibitem

\bibitem[\protect\citeauthoryear{Newman and Girvan}{2004}]{newman2004finding}
\begin{barticle}
\bauthor{\bsnm{Newman}, \binits{M.E.}},
\bauthor{\bsnm{Girvan}, \binits{M.}}:
\batitle{Finding and evaluating community structure in networks}.
\bjtitle{Physical review E}
\bvolume{69}(\bissue{2}),
\bfpage{026113}
(\byear{2004})
\end{barticle}
\endbibitem

\bibitem[\protect\citeauthoryear{Newman}{2002}]{newman2002assortative}
\begin{barticle}
\bauthor{\bsnm{Newman}, \binits{M.E.}}:
\batitle{Assortative mixing in networks}.
\bjtitle{Physical review letters}
\bvolume{89}(\bissue{20}),
\bfpage{208701}
(\byear{2002})
\end{barticle}
\endbibitem

\bibitem[\protect\citeauthoryear{Bae and Kim}{2014}]{bae2014identifying}
\begin{barticle}
\bauthor{\bsnm{Bae}, \binits{J.}},
\bauthor{\bsnm{Kim}, \binits{S.}}:
\batitle{Identifying and ranking influential spreaders in complex networks by neighborhood coreness}.
\bjtitle{Physica A: Statistical Mechanics and its Applications}
\bvolume{395},
\bfpage{549}--\blpage{559}
(\byear{2014})
\end{barticle}
\endbibitem

\end{thebibliography}

\section*{Supplementary information}
\subsection*{SI.1 Supplementary Methods}\label{subsec-Methods}

\begin{enumerate}[1.]
	\item[1.] \textbf{Curriculum reform time}
\end{enumerate}

Since 2000, Chinese National College Entrance Examination has undergone four rounds of curriculum reform, including the pilot draft of the curriculum plan and standards for ordinary high schools in 2003, the National Medium- and Long-Term Education Reform and Development Plan (2010-2020) in 2010, the new Gaokao policy in 2014, and the newly revised physics curriculum standards in 2017. These reforms have significantly influenced the college entrance examination. To further investigate the impact of these reforms on the physics examinations, we categorized 35 examinations into four stages based on the release times of the aforementioned policies: the first stage (from 2006 to 2010), the second stage (from 2011 to 2014), the third stage (from 2015 to 2017) and the fourth stage (from 2018 to 2020). For examinations in each stage, we constructed the corresponding IKPN and analyzed its topological characteristics using network analysis to explore the evolutionary trends of NCEE physics in response to these reforms.

\begin{enumerate}[2.]
	\item[2.] \textbf{Different volumes}
\end{enumerate}

As previously stated, the examination papers under investigation stem from the Chinese National College Entrance Examination system. These papers, referred to as ``National Papers'', are crafted by the National Education Examination Authority and are employed across the majority of provinces nationwide for the college entrance examination, with the aim of ensuring fairness in talent selection. The National Papers are categorized into three types: National Volume 1, National Volume 2 and National Volume 3, denoted as V1, V2 and V3, respectively. The propositions for these papers are formulated based on the examination outline and prescribed textbook materials. Typically, the ranking order based on difficulty, from highest to lowest, is V1, V2, and V3, respectively. Do these various types exhibit similar characteristics when forming networks? To explore this inquiry, we classified 35 papers according to their volume types. Among these, the V1 category comprises 15 papers, as does the V2 category. However, it is noteworthy that V3, having released only 5 papers in 2016, is excluded from analysis to mitigate potential impact discrepancies.

\begin{enumerate}[3.]
	\item[3.] \textbf{Different questions}
\end{enumerate}

The question types of the college entrance examinations of physics comprise choice questions, experimental questions, calculation questions and elective questions. To investigate the inherent knowledge structure characteristics and topological properties of these question types, we categorized and synthesized questions from 35 examinations based on their respective characteristics. This process yielded a set of questions corresponding to each question type. Subsequently, employing the aforementioned network construction method, we constructed IKPN based on question sets. Employing network analysis techniques, we thoroughly examined the network structural characteristics of different question types.

\begin{enumerate}[4.]
	\item[4.] \textbf{High school physics}
\end{enumerate}

High school physics courses have an extensive curriculum, including topics such as mechanics, thermodynamics, optics, electromagnetics and atomic physics. Among these, mechanics and electromagnetics are often considered particularly important subjects. Within mechanics, the subfields of statics, kinematics and dynamics explore the laws governing the static properties, dynamic properties and forces of objects, including the principles governing interactions among objects in the natural world. Electromagnetics includes the fundamental principles of electricity and magnetism, offering theoretical foundations for comprehending electromagnetic phenomena and the propagation of electromagnetic waves.

\begin{enumerate}[5.]
	\item[5.] \textbf{Examination data processing}
\end{enumerate}

When extracting key concepts, first consider the laws and theorems relevant to the question. Unless specifically emphasized by the question, do not extract the smaller concepts contained within these laws or theorems. For instance, if a question examines Newton's second law, which inherently involves concepts such as acceleration and force, extract "Newton's second law" and "Acceleration" if the question explicitly emphasizes calculating acceleration. Otherwise, only extract "Newton's second law."  Figure.~\ref{fig_question} illustrates an example of how a specific question is divided into key concepts.

\subsection*{SI.2 Supplementary  Figures}\label{subsec-Figures}

\setcounter{figure}{0}

\begin{figure}[H]
	\centering
	\includegraphics[width=1\textwidth]{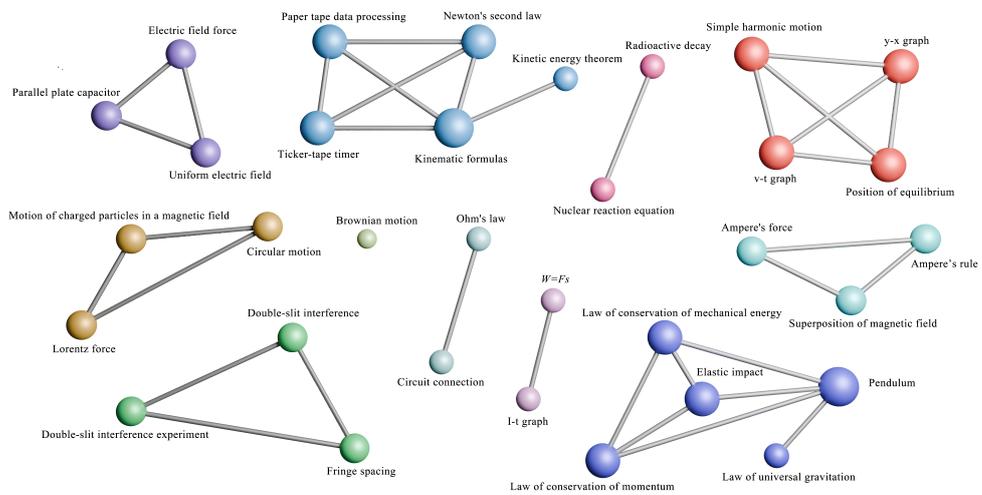}
	\caption{\textbf{The KPN of 12V2 is divided into different colors according to the community.} The KPN primarily consists of three dual-node communities, four triangular communities, three complex communities, and one isolated node community.}
	\label{fig:1}
\end{figure}

\begin{figure}[H]
	\centering
	\includegraphics[width=0.9\textwidth]{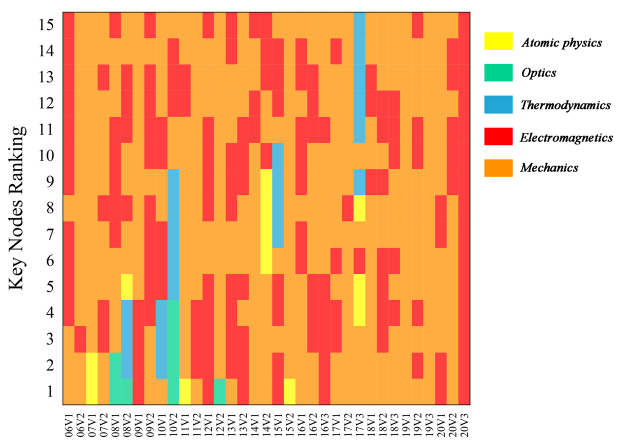}
	\caption{\textbf{The distribution diagram of the top 15 key knowledge points among the 35 KPNs.} In this diagram, orange represents knowledge points related to mechanics, red represents knowledge points related to electromagnetics, blue represents knowledge points related to thermodynamics, green represents knowledge points related to optics, and yellow represents knowledge points related to atomic physics. The ranking of key nodes is based on eigenvector centrality ($E$), with the diagram being sorted in descending order. Thus, the 15th position on the vertical axis corresponds to the node with the highest eigenvector centrality, ranked first.}
	\label{fig:2}
\end{figure}

\begin{figure}[H]
	\centering
	\includegraphics[width=0.9\textwidth]{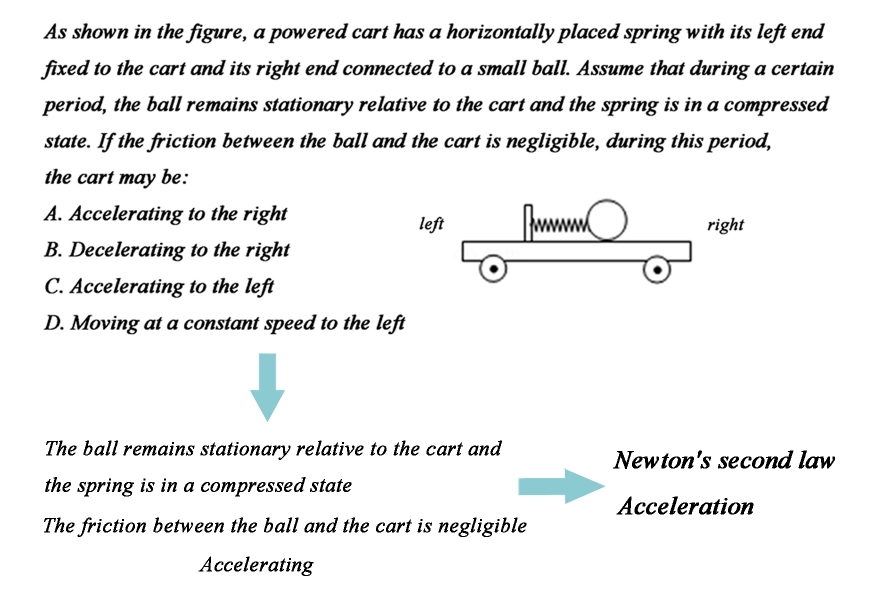}
	\caption{\textbf{An example of how a specific question is divided into key concepts.} The selected examination is 08V1, question two. Firstly, crucial information was extracted from the question statement, specifically "the ball remains stationary relative to the cart and the spring is in a compressed state," "the friction between the ball and the cart is negligible," and from the options, "Accelerating to the," . Secondly, based on these key pieces of information, the underlying concepts were identified, revealing that the question examines Newton's second law. Additionally, given the emphasis on "Accelerating" in the options, the final identified key concepts for this problem are "Newton's second law" and "Acceleration".}
	\label{fig_question}
\end{figure}

\subsection*{SI.3 Supplementary Tables}\label{subsec-Tables}

\setcounter{table}{0}

\begin{table}[h]
	\caption{\textbf{Main statistical characteristics of 35 knowledge point networks (KPNs).} Different volumes in various years are abbreviated; for example, 06V1 represents the KPN of National Volume 1 from 2006. \# of nodes represents the number of nodes in the KPN. \# of edges represents the number of edges in the KPN.} \label{tab1}%
	\begin{tabular}{@{}llllllll@{}}
		\toprule
		\textbf{Topological quantities} & \textbf{06V1} & \textbf{06V2} & \textbf{07V1} & \textbf{07V2} & \textbf{08V1} & \textbf{08V2} & \textbf{09V1} \\
		\midrule
		\# of nodes                     & 42            & 35            & 44            & 35            & 33            & 31            & 42            \\
		\# of edges                     & 70            & 51            & 73            & 53            & 50            & 40            & 62            \\
		Diameter                        & 3             & 3             & 3             & 4             & 4             & 4             & 4             \\
		Density                         & 0.081         & 0.086         & 0.077         & 0.089         & 0.095         & 0.086         & 0.072         \\
		Assortativity                   & 0.466         & 0.502         & 0.479         & 0.159         & 0.430         & 0.498         & 0.518         \\
		Transitivity                    & 0.86          & 0.82          & 0.85          & 0.61          & 0.77          & 0.74          & 0.87          \\
		Average degree                  & 3.33          & 2.91          & 3.32          & 3.03          & 3.03          & 2.58          & 2.95          \\
		Average shortest path length    & 1.95          & 1.76          & 1.96          & 2.48          & 2.06          & 1.87          & 2.02          \\
		Average clustering coefficient  & 0.85          & 0.82          & 0.86          & 0.85          & 0.36          & 0.39          & 0.86          \\ \midrule
		\textbf{Topological quantities} & \textbf{09V2} & \textbf{10V1} & \textbf{10V2} & \textbf{11V1} & \textbf{11V2} & \textbf{12V1} & \textbf{12V2} \\
		\# of nodes                     & 46            & 27            & 32            & 47            & 28            & 39            & 33            \\
		\# of edges                     & 87            & 21            & 38            & 55            & 36            & 40            & 35            \\
		Diameter                        & 4             & 2             & 2             & 3             & 3             & 2             & 2             \\
		Density                         & 0.084         & 0.060         & 0.077         & 0.051         & 0.095         & 0.054         & 0.066         \\
		Assortativity                   & 0.322         & 0.232         & 0.747         & 0.302         & -0.091        & 0.521         & 0.514         \\
		Transitivity                    & 0.73          & 0.79          & 0.93          & 0.77          & 0.76          & 0.91          & 0.89          \\
		Average degree                  & 3.78          & 1.56          & 2.38          & 2.34          & 2.57          & 2.05          & 2.12          \\
		Average shortest path length    & 2.22          & 1.33          & 1.27          & 1.64          & 1.62          & 1.47          & 1.30          \\
		Average clustering coefficient  & 0.86          & 0.51          & 0.59          & 0.80          & 0.35          & 0.57          & 0.70          \\ \midrule
		\textbf{Topological quantities} & \textbf{13V1} & \textbf{13V2} & \textbf{14V1} & \textbf{14V2} & \textbf{15V1} & \textbf{15V2} & \textbf{16V1} \\
		\midrule
		\# of nodes                     & 80            & 46            & 43            & 47            & 50            & 40            & 51            \\
		\# of edges                     & 232           & 72            & 53            & 61            & 73            & 45            & 91            \\
		Diameter                        & 5             & 4             & 3             & 3             & 3             & 5             & 4             \\
		Density                         & 0.073         & 0.070         & 0.059         & 0.056         & 0.060         & 0.058         & 0.071         \\
		Assortativity                   & 0.379         & 0.081         & 0.270         & 0.406         & 0.340         & 0.558         & 0.353         \\
		Transitivity                    & 0.74          & 0.63          & 0.71          & 0.86          & 0.82          & 0.83          & 0.73          \\
		Average degree                  & 5.80          & 3.13          & 2.47          & 2.60          & 2.92          & 2.25          & 3.57          \\
		Average shortest path length    & 2.66          & 2.28          & 1.97          & 2.20          & 1.62          & 2.47          & 2.06          \\
		Average clustering coefficient  & 0.46          & 0.38          & 0.78          & 0.36          & 0.81          & 0.35          & 0.41          \\ \midrule
		\textbf{Topological quantities} & \textbf{16V2} & \textbf{16V3} & \textbf{17V1} & \textbf{17V2} & \textbf{17V3} & \textbf{18V1} & \textbf{18V2} \\
		\midrule
		\# of nodes                     & 41            & 41            & 42            & 39            & 45            & 47            & 46            \\
		\# of edges                     & 65            & 41            & 69            & 72            & 63            & 85            & 66            \\
		Diameter                        & 5             & 2             & 3             & 5             & 1             & 6             & 3             \\
		Density                         & 0.079         & 0.050         & 0.080         & 0.097         & 0.064         & 0.079         & 0.064         \\
		Assortativity                   & 0.123         & 0.519         & 0.551         & 0.546         & 0.940         & 0.546         & 0.755         \\
		Transitivity                    & 0.65          & 0.89          & 0.75          & 0.75          & 0.97          & 0.75          & 0.92          \\
		Average degree                  & 3.17          & 2.00          & 3.29          & 3.69          & 2.80          & 3.62          & 2.87          \\
		Average shortest path length    & 2.61          & 2.00          & 2.00          & 2.43          & 1.00          & 2.73          & 1.39          \\
		Average clustering coefficient  & 0.33          & 0.23          & 0.72          & 0.34          & 0.73          & 0.80          & 0.37          \\ \midrule
		\textbf{Topological quantities} & \textbf{18V3} & \textbf{19V1} & \textbf{19V2} & \textbf{19V3} & \textbf{20V1} & \textbf{20V2} & \textbf{20V3} \\
		\midrule
		\# of nodes                     & 37            & 56            & 43            & 55            & 40            & 45            & 62            \\
		\# of edges                     & 59            & 151           & 79            & 141           & 70            & 72            & 140           \\
		Diameter                        & 3             & 7             & 5             & 5             & 3             & 4             & 3             \\
		Density                         & 0.089         & 0.098         & 0.087         & 0.095         & 0.090         & 0.073         & 0.074         \\
		Assortativity                   & 0.043         & 0.293         & 0.538         & 0.037         & 0.090         & 0.428         & 0.353         \\
		Transitivity                    & 0.63          & 0.75          & 0.67          & 0.75          & 0.68          & 0.73          & 0.90          \\
		Average degree                  & 3.19          & 5.39          & 3.67          & 5.13          & 3.50          & 3.20          & 4.52          \\
		Average shortest path length    & 1.83          & 3.01          & 2.41          & 2.64          & 1.91          & 2.21          & 1.70          \\
		Average clustering coefficient  & 0.35          & 0.42          & 0.38          & 0.31          & 0.65          & 0.37          & 0.45          \\
		\bottomrule
	\end{tabular}
\end{table}

\begin{table}[h]
	\caption{\textbf{Ranked among the top 7 communities in the IKPN.} Here N represents the number of knowledge points in each community, and subject represents the physical module corresponding to the community.}	\label{tab3}
	\begin{tabular}{@{}lll@{}}
		\toprule
		Rank & N   & Subject                                      \\
		\midrule
		1    & 172 & Mechanics, Electromagnetics, Physical optics \\
		2    & 27  & Thermodynamics                               \\
		3    & 20  & Mechanical vibration and mechanical waves    \\
		4    & 14  & Geometrical optics                           \\
		5    & 13  & Atomic physics                               \\
		6    & 11  & Electrical experiment                        \\
		7    & 9   & Circuit and current                          \\
		\bottomrule
	\end{tabular}
\end{table}

\begin{table}[h]
	\caption{\textbf{Top 30 key knowledge points of the IKPN.} Eigenvector Centrality represents the eigenvector centrality $E$ index. The top 30 values are extracted and ranked in descending order according to their centrality index.}	\label{tab4}
	\begin{tabular}{@{}lll@{}}
		\toprule
		Rank & Node                                             & Eigenvector Centrality \\
		\midrule
		1    & Newton's second law                              & 0.423                  \\
		2    & Circular motion                                  & 0.386                  \\
		3    & Kinematic formulas                               & 0.311                  \\
		4    & Kinetic energy theorem                           & 0.309                  \\
		5    & Lorentz force                                    & 0.308                  \\
		6    & Motion of charged particles in a magnetic field  & 0.287                  \\
		7    & Motion of charged particles in an electric field & 0.216                  \\
		8    & Uniform variable rectilinear motion              & 0.176                  \\
		9    & Law of conservation of mechanical energy         & 0.159                  \\
		10   & Law of conservation of momentum                  & 0.133                  \\
		11   & Centripetal force                                & 0.126                  \\
		12   & Horizontal projectile motion                     & 0.107                  \\
		13   & Law of universal gravitation                     & 0.102                  \\
		14   & Electric field strength                          & 0.092                  \\
		15   & Electric potential                               & 0.092                  \\
		16   & Law of conservation of energy                    & 0.077                  \\
		17   & Theorem of momentum                              & 0.077                  \\
		18   & Law of electromagnetic induction                 & 0.077                  \\
		19   & Work-energy principle                            & 0.075                  \\
		20   & Electric field force                             & 0.073                  \\
		21   & Ampere's force                                   & 0.071                  \\
		22   & v-t graph                                        & 0.070                  \\
		23   & Class of flat parabolic motion                   & 0.066                  \\
		24   & Kinetic energy                                   & 0.065                  \\
		25   & Uniform magnetic field                           & 0.064                  \\
		26   & Acceleration                                     & 0.063                  \\
		27   & Left-hand rule                                   & 0.062                  \\
		28   & Uniform electric field                           & 0.062                  \\
		29   & Dynamic friction factor                          & 0.061                  \\
		30   & Ohm's law                                        & 0.060                  \\
		\bottomrule
	\end{tabular}
\end{table}

\begin{table}[h]
	\caption{\textbf{Key knowledge points in two volumes.} The top 15 nodes of the two KPNs were selected based on the Eigenvector Centrality $E$.}\label{tab5}
	\begin{tabular}{@{}ll@{}}
		\toprule
		V1                                           & V2                                           \\
		\midrule
		Newton's second law                          & Newton's second law                          \\
		Circular motion                              & Kinematic formulas                           \\
		Motion of charged particle in magnetic field & Circular motion                              \\
		Lorentz force                                & Lorentz force                                \\
		Motion of charged particle in electric field & Motion of charged particle in magnetic field \\
		Kinetic energy theorem                       & Kinetic energy theorem                       \\
		Kinematic formulas                           & Uniform variable rectilinear motion          \\
		Uniform variable rectilinear motion          & Law of conservation of mechanical energy     \\
		Law of conservation of momentum              & Motion of charged particle in electric field \\
		Law of conservation of mechanical energy     & Horizontal projectile motion                 \\
		Electric field strength                      & Electric field force                         \\
		Law of universal gravitation                 & Electric potential                           \\
		Acceleration                                 & Uniform electric field                       \\
		Centripetal force                            & Law of conservation of momentum              \\
		Electric potential                           & Centripetal force                            \\
		\bottomrule
	\end{tabular}
\end{table}

\begin{table}[h]
	\caption{\textbf{Comparison of V1 and V2 in four types of question networks.} The four types of questions are categorized based on V1 and V2, respectively, to construct V1 and V2 KPN based on question types.}\label{tab6}
	\begin{tabular}{@{}lllllllll@{}}
		\toprule
		\multirow{2}*{Property}        & \multicolumn{2}{l}{Choice} & \multicolumn{2}{l}{Experiment} & \multicolumn{2}{l}{Calculation} & \multicolumn{2}{l}{Elective}                                 \\
		                               & V1                         & V2                             & V1                              & V2                           & V1    & V2    & V1    & V2    \\
		\midrule
		\# of nodes                    & 168                        & 153                            & 51                              & 46                           & 70    & 60    & 79    & 68    \\
		\# of edges                    & 524                        & 362                            & 73                              & 66                           & 254   & 214   & 160   & 98    \\
		Density                        & 0.037                      & 0.031                          & 0.057                           & 0.064                        & 0.105 & 0.121 & 0.052 & 0.043 \\
		Transitivity                   & 0.45                       & 0.37                           & 0.45                            & 0.47                         & 0.48  & 0.39  & 0.58  & 0.71  \\
		Average degree                 & 6.24                       & 4.73                           & 2.86                            & 2.87                         & 7.26  & 7.13  & 4.05  & 2.88  \\
		Average clustering coefficient & 0.16                       & 0.09                           & 0.26                            & 0.22                         & 0.15  & 0.13  & 0.19  & 0.20  \\
		$F_d$(x10)                     & 0.16                       & 0.05                           & 0.19                            & 0.19                         & 0.54  & 0.42  & 0.23  & 0.17  \\
		\bottomrule
	\end{tabular}
\end{table}

\begin{table}[h]
	\caption{ \textbf{Key knowledge points of networks in four question types.} Based on Eigenvector Centrality$E$, the top 15 nodes for each of the four types of questions were selected.}	\label{tab7}
	\begin{tabular}{p{4cm} p{3cm} p{4cm} p{3cm}}
		\toprule
		Choice                                           & Experiment                     & Calculation                                      & Elective                  \\
		\midrule
		Circular motion                                  & Ohm's law                      & Newton's second law                              & Snell's law               \\
		Lorentz force                                    & Circuit connection             & Kinematic formulas                               & Total internal reflection \\
		Newton's second law                              & Electricity meter modification & Kinetic energy theorem                           & Critical angle            \\
		Motion of charged particles in a magnetic field  & Thermistor                     & Circular motion                                  & Index of refraction       \\
		Kinetic energy theorem                           & Multimeter                     & Motion of charged particles in a magnetic field  & Reflection law            \\
		Electric potential                               & Voltammetry                    & Lorentz force                                    & Lightspeed                \\
		Law of universal gravitation                     & Internal resistance            & Motion of charged particles in an electric field & Angle of refraction       \\
		Centripetal force                                & Slide rheostat                 & Uniform variable rectilinear motion              & Frequency                 \\
		Motion of charged particles in an electric field & Resistance                     & Law of conservation of momentum                  & Optical path diagram      \\
		Electric potential energy                        & Circuit design                 & Law of conservation of mechanical energy         & Angle of reflection       \\
		Left-hand rule                                   & Semi-deflection method         & Theorem of momentum                              & Angle of incidence        \\
		Electric field strength                          & Voltage                        & Horizontal projectile motion                     & Refraction of light       \\
		Law of electromagnetic induction                 & Electric current               & Class of flat parabolic motion                   & Wave speed                \\
		Kinetic energy                                   & Resistance box reading         & Electric field strength                          & Period                    \\
		Law of conservation of mechanical energy         & Series voltage division        & Uniform electric field                           & Wavelength                \\
		\bottomrule
	\end{tabular}
\end{table}

\begin{table}[h]
	\caption{\textbf{The correlation between the small-world effect and exam difficulty.} N represents the sample size used for calculating correlations (KPNs exhibiting small-world properties). The Spearman correlation coefficient indicates the correlation derived using Spearman's rank correlation. Spearman's correlation was chosen because some data did not follow a normal distribution.The P-value represents the significance level. ALL refers to all 30 KPNs exhibiting small-world properties, while V1 denotes the 14 KPNs categorized as V1 type that exhibit small-world properties.}	\label{tab10}
	\begin{tabular}{@{}llll@{}}
		\toprule
		    & N  & Spearman correlation coefficient & P-value               \\
		\midrule
		ALL & 30 & 0.6080                           & $3.65 \times 10^{-4}$ \\
		V1  & 14 & 0.5516                           & 0.040849058           \\
		V2  & 13 & 0.6319                           & 0.020516043           \\
		\bottomrule
	\end{tabular}
\end{table}

\begin{table}[h]
	\caption{\textbf{Explanation about some topological indicators of complex network.} All network metrics mentioned in this article are introduced here.}\label{tab9}
	\begin{tabular}{p{5cm} p{10.5cm}}
		\toprule
		Indicators                                                     & Implications                                                                                                                                                                                                                                                                                                                                                               \\ 
		\midrule
		Degree $k_i$                                                   & The degree $k_i$ is the sum of the edge weights for edges connected to node $i$.                                                                                                                                                                                                                                                                                           \\
		Average degree $\left\langle k \right\rangle$                  & Signifying the average number of connections per node in a network. A higher average degree implies denser connections between nodes.                                                                                                                                                                                                                                      \\
		Degree distribution $P(k)$                                     & Giving the probability that a node chosen uniformly at random has degree $k$. A network whose node degree distribution obeys the power law distribution characteristics is called a scale-free network\cite{barabasi1999emergence}.                                                                                                                                        \\
		Diameter $d$                                                   & The longest path length between node pairs in the network.                                                                                                                                                                                                                                                                                                                 \\
		Density $\rho$                                                 & Represents the ratio between the actual number of existing edges and the maximum potential number of edges. It quantifies the actual level of connections within the network.                                                                                                                                                                                              \\
		Modularity $Q$                                                 & Modularity is a measure to evaluate the effectiveness of community detection\cite{newman2004finding}.                                                                                                                                                                                                                                                                      \\
		Transitivity $T$                                               & Calculating the score of all possible triangles present within the network. Networks with higher transitivity typically exhibit more closed triads and a higher degree of clustering among nodes.                                                                                                                                                                          \\
		Assortativity $a$                                              & A network is said to show assortative mixing if the nodes in the network that have many connections tend to be connected to other nodes with many connections. Degree correlation is a measure of assortative mixing for networks, which can be determined from the Pearson correlation coefficient of the degrees at both ends of the edges \cite{newman2002assortative}. \\
		Average shortest path length $\left\langle \ell \right\rangle$ & Refers to the average of the shortest path between all pairs of nodes. This metric is associated with information propagation and connectivity within the network.                                                                                                                                                                                                         \\
		Average clustering coefficient $\left\langle c \right\rangle$  & The clustering coefficient $c_i$ is calculated as the number of weighted connections between the neighbors of node $i$ divided by the maximum possible number of weighted connections among these neighbors, and the average clustering coefficient is the average of clustering coefficients of all nodes in the network.                                                 \\
		Degree centrality $D$                                          & Which is defined as the ratio of connections incident upon a node. The higher the degree of a node, the higher its degree centrality.                                                                                                                                                                                                                                      \\
		Eigenvector centrality $E$                                     & Which is a measure of the influence of a node in a connected network. A high eigenvector score means that a node is connected to many nodes who themselves have high scores. That is, the more important a node?s neighbors are, the more important the node is.                                                                                                           \\
		Closeness centrality $C$                                       & Closeness centrality of a node calculated as the reciprocal of the sum of the length of the shortest paths between the node and all other nodes in the network, which reflects the proximity between a node and other nodes in the network.                                                                                                                                \\
		Betweenness centrality $B$                                     & Which is a measure of centrality in a graph based on shortest paths. It uses the number of shortest paths passing through a node to characterize the importance of the node.                                                                                                                                                                                               \\
		Monotonicity $M$                                               & Which calculates the discriminating effect of various algorithm indicators on node importance\cite{bae2014identifying}.                                                                                                                                                                                                                                                    \\
		\bottomrule
	\end{tabular}
\end{table}

\begin{table}[h]
	\caption{\textbf{Average scores and network indicators for the NCEE physics exams in four provinces.} For each province, data from two different years were selected for comparison. $F_d$ represents the comprehensive difficulty coefficient. This metric is multiplied by 10 to output more reasonable numerical values for enhanced readability and interpretability.}	\label{tab10}
	\begin{tabular}{@{}lllllllll@{}}
		\toprule
		Properties                     & SX    & SX    & HN    & HN    & HB    & HB    & FJ    & FJ   \\
		\midrule
		Year                           & 2017  & 2018  & 2020  & 2021  & 2021  & 2022  & 2022  & 2023 \\
		\# of nodes                    & 39    & 46    & 47    & 49    & 54    & 64    & 36    & 39   \\
		\# of edges                    & 72    & 66    & 67    & 86    & 119   & 191   & 61    & 74   \\
		Density                        & 0.097 & 0.064 & 0.062 & 0.073 & 0.083 & 0.095 & 0.097 & 0.10 \\
		Transitivity                   & 0.75  & 0.92  & 0.78  & 0.70  & 0.70  & 0.66  & 0.57  & 0.62 \\
		Average degree                 & 3.69  & 2.87  & 2.85  & 3.51  & 4.41  & 5.97  & 3.39  & 3.79 \\
		Average clustering coefficient & 0.34  & 0.37  & 0.36  & 0.46  & 0.43  & 0.46  & 0.38  & 0.43 \\
		Average score                  & 42.2  & 48.8  & 46    & 44.5  & 58.1  & 52    & 56.5  & 51.2 \\
		$F_d$(x10)                     & 0.92  & 0.62  & 0.50  & 0.82  & 1.09  & 1.70  & 0.72  & 1.01 \\
		\bottomrule
	\end{tabular}
\end{table}

\end{document}